\documentclass[prx,letterpaper,nobalancelastpage,twocolumn,superscriptaddress,nofootinbib]{revtex4-2}

\usepackage{graphicx}
\usepackage{amsmath}
\usepackage{bbold}
\usepackage{amssymb}
\usepackage[english]{babel}
\usepackage{color}
\usepackage[version=4]{mhchem}
\usepackage[hidelinks]{hyperref}
\usepackage{dsfont}
\usepackage{placeins}
\usepackage{mathtools}
\usepackage[normalem]{ulem}

\setcitestyle{super}

\usepackage{soul}
\newcommand{\ket}[1]{| #1 \rangle}
\newcommand{\bra}[1]{\langle #1 |}

\newcommand{\tr}[1]{\mathrm{tr}\left\{#1\right\}}

\newcommand{\Caltech}{California Institute of Technology, Pasadena, CA 91125, USA}

\newcommand{\Berkeley}{Department of Physics, University of California, Berkeley, CA 94720, USA}
\newcommand{\UIUC}{Department of Physics, The University of Illinois at Urbana-Champaign, Urbana, Illinois 61801-3080, USA}
\newcommand{\MIT}{Center for Theoretical Physics, Massachusetts Institute of Technology, Cambridge, MA 02139, USA}
\newcommand{\Harvard}{Harvard University, Cambridge, MA 02138, USA}

\newcommand{\Innsbruck}{Institute for Theoretical Physics, University of Innsbruck, Innsbruck A-6020, Austria}
\newcommand{\InnsbruckCenter}{Institute for Quantum Optics and Quantum Information, Austrian Academy of Sciences, Innsbruck A-6020}

\usepackage{caption}

\DeclareCaptionLabelSeparator{bar}{ \textbf{\textbar}~}
\usepackage{enumitem}
\setlist{nolistsep}

\begin{document}

\title{Preparing random states and benchmarking with many-body quantum chaos}
\author{Joonhee Choi}\thanks{These authors contributed equally to this work}
\author{Adam L. Shaw}\thanks{These authors contributed equally to this work}
\author{Ivaylo S. Madjarov}
\author{Xin Xie}
\author{Ran Finkelstein}
\affiliation{\Caltech}
\author{\\Jacob P. Covey}
\affiliation{\Caltech}
\affiliation{\UIUC}
\author{Jordan S. Cotler}
\affiliation{\Harvard}
\author{Daniel K. Mark}
\affiliation{\MIT}
\author{Hsin-Yuan Huang}
\affiliation{\Caltech}
\author{Anant Kale}
\affiliation{\Harvard}
\author{\\Hannes Pichler}
\affiliation{\Innsbruck}
\affiliation{\InnsbruckCenter}
\author{Fernando G.S.L. Brand{\~{a}}o}
\affiliation{\Caltech}
\author{Soonwon Choi}\email{soonwon@mit.edu}
\affiliation{\MIT}
\affiliation{\Berkeley}
\author{Manuel Endres}\email{mendres@caltech.edu}
\affiliation{\Caltech}

\maketitle

\textbf{Producing quantum states at random has become increasingly important in modern quantum science, with applications both theoretical and practical. In particular, ensembles of such randomly-distributed, but pure, quantum states underly our understanding of complexity in quantum circuits~\cite{Brandao2021} and black holes~\cite{Hayden2007}, and have been used for benchmarking quantum devices~\cite{Neill2018,Cross2019} in tests of quantum advantage~\cite{Arute2019,Wu2021}. However, creating random ensembles has necessitated a high degree of spatio-temporal control~\cite{Emerson2003,Harrow2009,Dankert2009,Brandao2016,Ohliger2013,Onorati2017,Nakata2017,Elben2018}, placing such studies out of reach for a wide class of quantum systems. Here we solve this problem by predicting and experimentally observing the emergence of random state ensembles naturally under time-independent Hamiltonian dynamics, which we use to implement an efficient, widely applicable benchmarking protocol. The observed random ensembles emerge from projective measurements and are intimately linked to universal correlations built up between subsystems of a larger quantum system, offering new insights into quantum thermalization~\cite{Kaufman2016,Popescu2006}. Predicated on this discovery, we develop a fidelity estimation scheme, which we demonstrate for a Rydberg quantum simulator with up to 25 atoms using fewer than $10^4$ experimental samples. This method has broad applicability, as we show for Hamiltonian parameter estimation, target-state generation benchmarking, and comparison of analog and digital quantum devices. Our work has implications for understanding randomness in quantum dynamics~\cite{Cotler2021}, and enables applications of this concept in a much wider context~\cite{Brandao2016,Boixo2018,Arute2019,Bouland2019,Haferkamp2020,Dankert2009,Boixo2018,Cross2019,Brydges2019,Elben2020,Huang2020}.} 

\begin{figure}[t!]
	\centering
	\includegraphics[width=89mm]{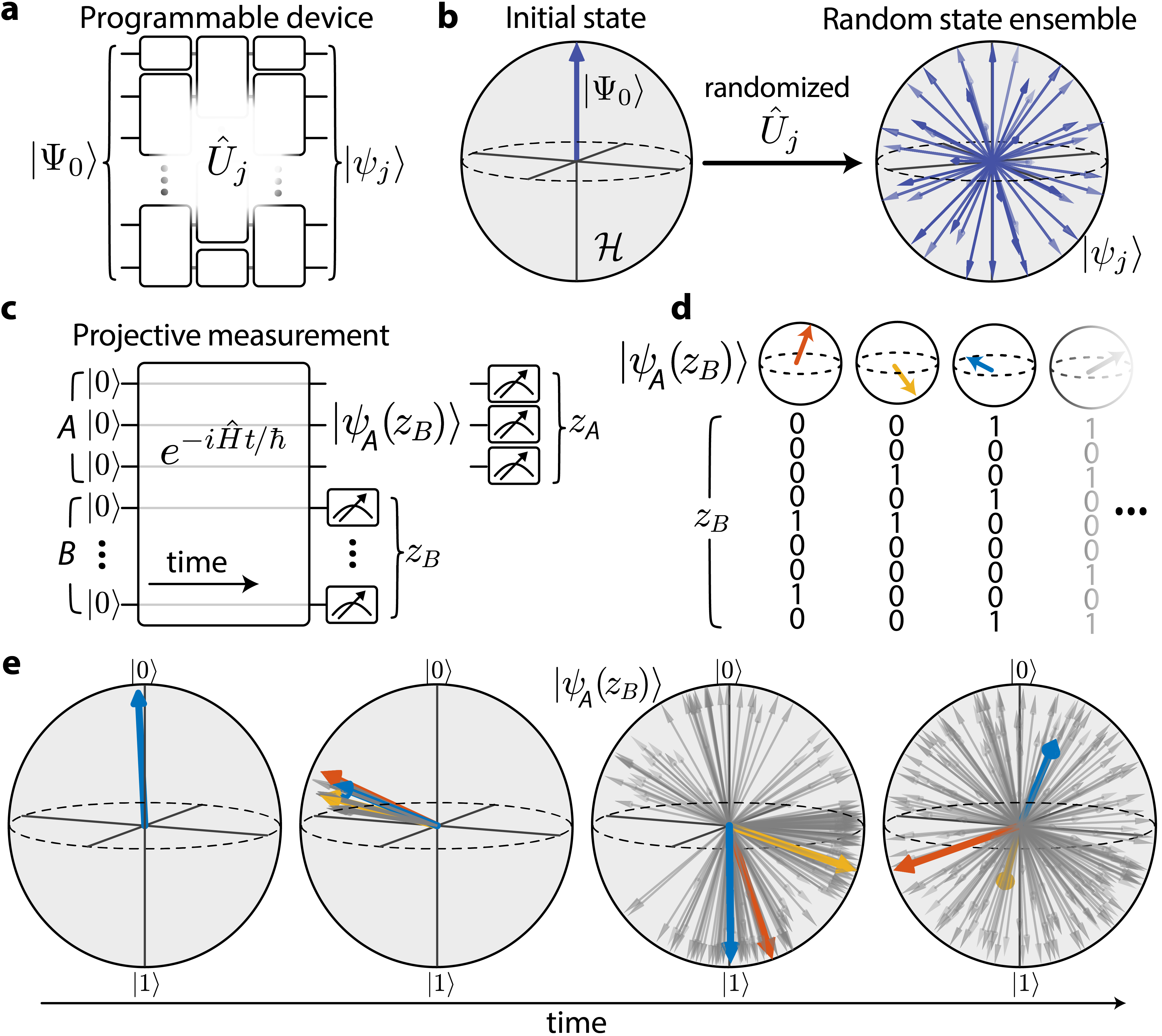}
	\caption{\textbf{Random pure state ensembles from Hamiltonian dynamics. a,} A thought experiment, consisting of a programmable device producing arbitrary quantum states $\ket{\psi_j}$ through unitary operations $\hat{U}_j$, where $j$ enumerates over different program setting. \textbf{b,} Repeatedly applying explicitly randomized unitary evolution to an initial state $\ket{\Psi_0}$ produces an ensemble of pure quantum states $\ket{\psi_j}$ (blue arrows) which is distributed near-uniformly over the Hilbert space, $\mathcal{H}$ (grey sphere), a \textit{random state ensemble}. \textbf{c,} Here we demonstrate a new approach to creating random state ensembles based on only a single instance of time-independent Hamiltonian evolution. An initial product state evolves under a Hamiltonian, $\hat{H}$, before site-resolved projective measurement in the computational basis \{$\ket{0}$, $\ket{1}$\}. We bipartition the system into two subsystems $A$ and $B$, and analyze the conditional measurement outcomes in subsystem $A$, $z_A$, given a specific result $z_B$ from the complement $B$. These outcomes are described by the \textit{projected ensemble}, a pure state ensemble in $A$, $\{\ket{\psi_A(z_B)}\}$, realized through measurement of $B$. \textbf{d,} As an example for when $A$ consists of a single qubit, conditional single-qubit quantum states $\ket{\psi_A(z_B)}$ are visualized on a Bloch sphere for all possible $z_B$ bitstrings. \textbf{e,} Numerical simulations of our experimental system show that the distribution of the conditional pure state ensemble in $A$ changes during evolution into a near-uniform form, with selected states highlighted to demonstrate their late-time divergence despite similar initial conditions.
	} 
	\vspace{-0.5cm}
	\label{Fig1}
\end{figure}

We start by illustrating the concept of pure random state ensembles via a thought experiment: consider a programmable quantum device which evolves an input state $\ket{\Psi_0}$ to an arbitrary output state $\ket{\psi_j}$, labeled by the program setting, $j$ (Fig.~\ref{Fig1}a). If the set of states $\ket{\psi_j}$ -- in the limit of many repetitions with different $j$ -- is homogeneously distributed over the output Hilbert space, it is termed a Haar-random (or uniform) state ensemble~\cite{Harrow2013}. A distribution of states close to the Haar-random one is shown for a single qubit in Fig.~\ref{Fig1}b (right).

Practically, approximations to Haar-random state ensembles are generated by certain quantum devices requiring explicit classical randomization, in the sense that output states $\ket{\psi_j}$ are produced by randomly chosen unitary evolution operators $\hat{U}_j$. Examples include random unitary circuits~\cite{Brandao2016,Harrow2009}, where each configuration $j$ is realized by a random choice of single and two-qubit gates, and stochastic evolution with a dynamically changing Hamiltonian~\cite{Nakata2017,Li2019}, $\hat{H}_j(t)$. In such systems, generation of approximate random state ensembles have been used for benchmarking of large-scale quantum devices~\cite{Neill2018,Cross2019}, including fidelity estimation as part of quantum advantage~\cite{Arute2019,Wu2021} and quantum volume tests~\cite{Jurcevic2021}. On a more fundamental level, random ensembles provide important insights into studies of complexity growth in quantum systems~\cite{Brandao2021} and understanding the quantum properties of black holes~\cite{Hayden2007,Piroli2020}.

However, it is currently unknown how to generate such random ensembles from the simplest form of quantum evolution, that governed by a fixed, time-independent Hamiltonian $\hat{H}$ which is not explicitly randomized, as is the case for dynamics of closed and unperturbed quantum systems. Here, by considering pure state ensembles generated during partial measurement of a larger quantum system (Fig.~\ref{Fig1}c), we show such random ensembles \textit{do} in fact emerge under such conditions. These emergent random ensembles enable applications such as device benchmarking, even in systems without explicit local, time-resolved control, which we demonstrate here experimentally using a Rydberg atom simulator~\cite{Bernien2017,Browaeys2020,Madjarov2020} with up to 25 atoms.

\vspace{0.25cm}
\noindent\textbf{Observation of emergent randomness}\newline
To study the emergence of random state ensembles, we consider Hamiltonian evolution that produces a global quantum state $\ket{\psi}$, which we here suppose describes a set of qubits with basis states $\ket{0}$ and $\ket{1}$. We bipartition the state into two subsystems: a local system of interest, $A$, and its complement $B$ (Fig. ~\ref{Fig1}c). Explicitly keeping track of measurement results in $B$, which are bitstrings of the form e.g. $z_B=100\cdots010$, provides a full description of the total system state as
\begin{align}
\ket{\psi}=\sum\limits_{z_B}\sqrt{p(z_B)}\ket{\psi_A(z_B)}\otimes\ket{z_B},
\label{eq:psi_projected}
\end{align}
where $p(z_B)$ is the probability of measuring a given bitstring in $B$, and $\ket{\psi_A(z_B)}$ is a pure quantum state in $A$ conditioned on the measurement outcome in $B$. Thus, for each possible $z_B$, there is a well-defined pure state in $A$, the set of all of which is generally not orthogonal. Together these states, $\ket{\psi_A(z_B)}$, and their respective probabilities, $p(z_B)$, form what we term the \textit{projected ensemble}~\cite{Cotler2021} (Fig. ~\ref{Fig1}d); similar concepts also enter the definition of localizable entanglement~\cite{Verstraete2004,Popp2005}, and in the concept of conditional wavefunctions~\cite{Goldstein2006,Goldstein2016}. By tracking the time evolution of the projected ensemble through both the states and probabilities which compose it, we can probe for signatures of the ensemble approaching a Haar-random distribution (Fig.~\ref{Fig1}e).

We stress that this concept is distinct from typical studies of equilibration in quantum many-body systems. There, the central object of interest is the reduced density operator on $A$, $\hat{\rho}_A = \text{Tr}_B (\hat{\rho})$, found from tracing out $B$ from the full density operator $\hat{\rho}$. The reduced density operator can be constructed by averaging over the projected ensemble states, $\hat{\rho}_A=\sum_{z_B} p(z_B) \ket{\psi_A(z_B)}\bra{\psi_A(z_B)}$, but as such can only provide information on the mean of ensemble observables, and never on the actual ensemble distribution itself.

To elucidate the importance of this distinction and reveal the emergence of random statistical properties of the projected ensemble, we employ a Rydberg analog quantum simulator~\cite{Bernien2017,Browaeys2020,Madjarov2020}, implemented with alkaline-earth atoms~\cite{Norcia2018,Cooper2018,Saskin2019,Covey2019b}, which provides high fidelity preparation, evolution, and readout~\cite{Madjarov2020} (Ext. Data Fig.~\ref{EFig_feedback}, Methods). After a variable evolution time, we perform site-resolved readout in a fixed measurement basis, yielding experimentally measured bitstrings, $z$, which we bipartition into bitstrings $z_A$ and $z_B$ (Methods).

\begin{figure}[t!]
	\centering
	\includegraphics[width=89mm]{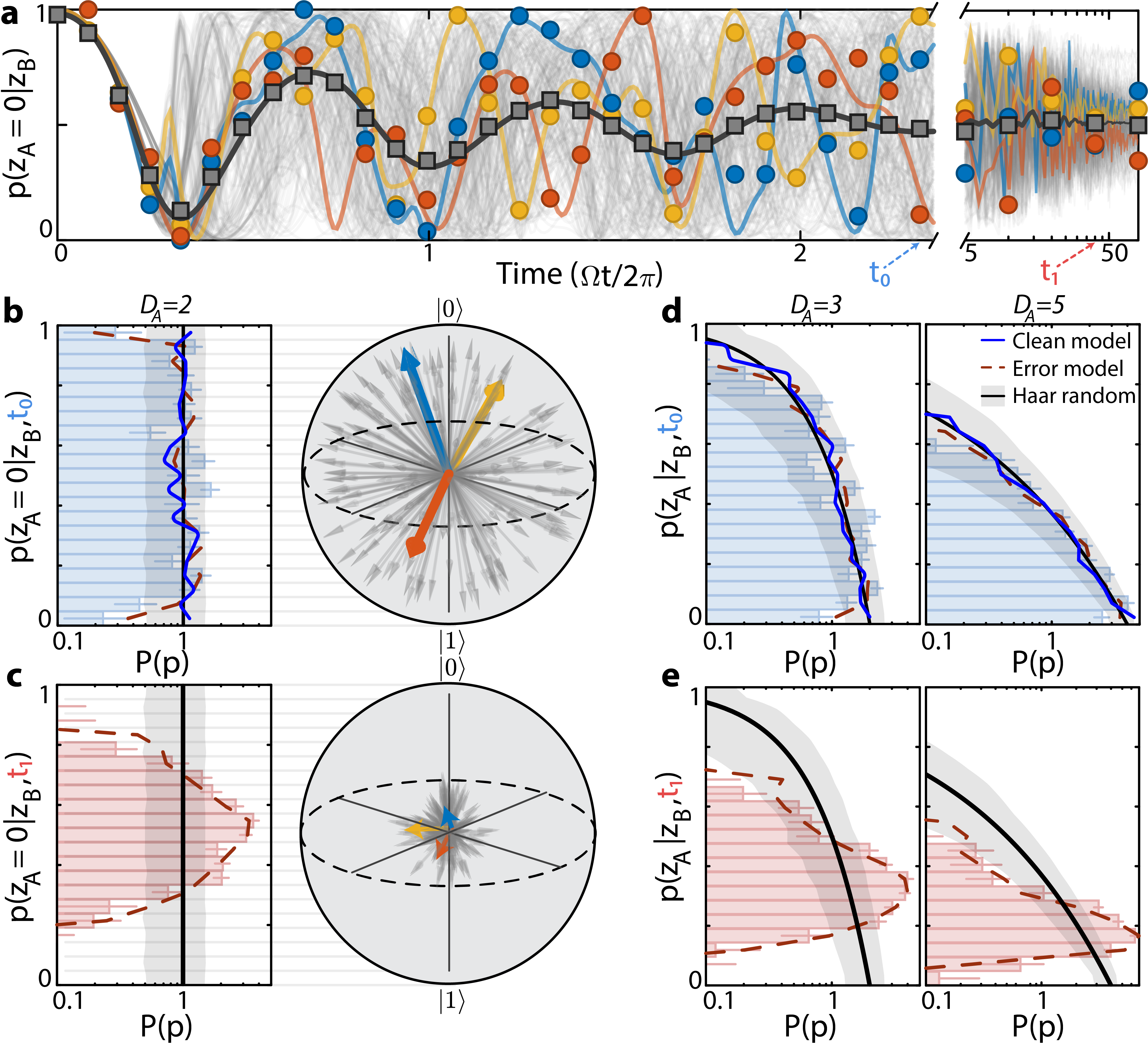}
	\caption{\textbf{Experimental signatures of random pure state ensembles. a,} We employ a ten-atom Rydberg quantum simulator (Ext. Data Fig.~\ref{EFig_feedback}) to perform Hamiltonian evolution leading to quantum thermalization at infinite effective temperature (see main text for details). For a single qubit in $A$, we plot the probabilities for finding a single qubit subsystem in state $\ket{0}$ as a function of time. Grey square markes indicate the \textit{marginal} probabilities $p(z_A{=}0)$, which equilibrate to ${\sim}0.5$ due to thermalization with $B$. In contrast, colored circle markers show \textit{conditional} probabilities given a specific measured $z_B$ in $B$, $p(z_A{=}0|z_B)$, which show large fluctuations even after the marginal probability reaches a steady state; these then diminish at late times due to extrinsic decoherence effects. Such conditional probabilities yield information about the projected ensemble as $p(z_A|z_B)=|\langle z_A|\psi_A(z_B)\rangle|^2$. Grey lines are simulated trajectories of $p(z_A{=}0|z_B)$ for all outcomes $z_B$, with a few highlighted to be compared with experimental data (color lines and markers). Decoherence sources~\cite{Supplement} are included for simulations after the axis break. \textbf{b,} Histograms, $P(p)$, of the probabilities $p(z_A{=}0|z_B)$ at intermediate ($\Omega t_0/2\pi = 2.3$) time. The experimental results are close to a flat distribution, consistent with a Haar-random ensemble, as visualized by the simulated distribution of projected states (right). \textbf{c,} However, at late ($\Omega t_1/2\pi = 38$) time, decoherence effects have concentrated probabilities around $1/D_A = 0.5$, consistent with the error model simulation showing the reduced lengths of single-qubit states (right). \textbf{d,e,} Similar agreement with predictions from random state ensembles is also seen for larger subsystem sizes of $A$ with higher subsystem dimension, $D_A$ (Methods). In \textbf{b-e}, black lines and grey bands are predictions and uncertainties (from finite sampling) of a $D_A$-dimensional uniform random ensemble; red dashed lines and blue solid lines are from simulations with and without decoherence~\cite{Supplement}, respectively.
	} 
	\vspace{-0.5cm}
	\label{Fig2}
\end{figure}

Hamiltonian parameters are chosen such that, after a short settling time, the \textit{marginal} probability, $p(z_A)$, of measuring a given $z_A$ (while ignoring the complementary $z_B$) agrees with the prediction from $\hat{\rho}_A$ being a maximally mixed state. In the language of quantum thermalization~\cite{Deutsch1991,Srednicki1994,Rigol2008,Nandkishore2015,Kaufman2016,Abanin2019,Ueda2020}, this prediction is equivalent to saying $\hat{\rho}_A$ has reached an equilibrium at infinite effective temperature with the complement $B$ as an effective, intrinsic bath~\cite{Kaufman2016,Popescu2006,delRio2016}. For a single qubit in $A$, such a reduced density operator is $\hat{\rho}_A = \frac{1}{2} \big( |0\rangle \langle 0| + | 1 \rangle \langle 1| \big)$: the qubit has a probability of being in state $\ket{0}$ of $p(z_A{=}0) = 1/D_A=1/2$, where $D_A=2$ is the local dimension of A. As shown in Fig.~\ref{Fig2}a, after a short transient period the experimentally measured probabilities, $p(z_A{=}0)$ (grey squares), equilibrate in agreement with this prediction. We note that post-selection is applied in accordance with the Rydberg blockade constraint (Methods).

We now contrast this equilibration with the dynamics of \textit{conditional} probabilities, $p(z_A|z_B)$, of measuring a given $z_A$ conditioned on finding an accompanying measurement outcome in the intrinsic bath, $z_B$. We note the marginal probability for finding $z_A$ is the weighted average over conditional probabilities, $p(z_A)=\sum_{z_B} p(z_B)p(z_A|z_B)$. More generally, while $p(z_A)$ yields information of the reduced density operator, such conditional probabilities yield signatures of the projected ensemble, as $p(z_A|z_B)=|\langle z_A|\psi_A(z_B)\rangle|^2$. In Fig.~\ref{Fig2}a, we plot numerically simulated $p(z_A{=}0|z_B)$ in grey, with selected traces highlighted in color to be compared with their corresponding experimental data (circle markers). Importantly, we find that the conditional probabilities are highly fluctuating in a seemingly chaotic fashion with sensitive dependence on $z_B$, even when the marginal probability has reached a steady state. In experiments, we note that these fluctuations slowly damp out over time due to extrinsic decoherence effects from coupling to an \textit{external} environment at very late time, but that these decoherence effects do not appear to affect the late-time marginal probability (right panel, Fig.~\ref{Fig2}a).

To analyze fluctuations quantitatively,  we construct a histogram $P(p)$ of finding the conditional probability $p(z_A|z_B)$ in an interval $[p,p+\Delta p]$, with $\Delta p$ the bin size (Fig.~\ref{Fig2}b). We plot such histograms for a time when fluctuations are strong and decoherence effects are small ($t_0$, Fig.~\ref{Fig2}b) as well as at very late time ($t_1$, Fig.~\ref{Fig2}c) when decoherence dominates. At $t_0$, the experimental $P(p)$ distribution is essentially flat, as predicted for a Haar-random ensemble, up to finite-sampling fluctuations and weak decoherence effects~\cite{Supplement}. We additionally show projected states obtained from simulation (Bloch sphere in Fig.~\ref{Fig2}b), including decoherence, to illustrate how such a flat distribution is generated from a near-uniform ensemble of states. At very late time, $t_1$, decoherence reduces the purity of projected states significantly, leading to $P(p)$ becoming concentrated around $1/D_A=0.5$ (Fig.~\ref{Fig2}c). This highlights that the agreement between the experimental data and the random ensemble prediction in Fig.~\ref{Fig2}b,d is a coherent phenomenon of closed quantum system dynamics. We further validate this in Figs.~\ref{Fig2}d,e by plotting the $P(p)$ for $A$ composed of 2 and 3 atoms, with corresponding Hilbert space dimensions of $D_A=3$ and $5$, respectively (Methods). Here, the prediction from the Haar-random distribution~\cite{Arute2019} is $P(p)=(D_A-1)(1-p)^{D_A-2}$, which we note in the limit $D_A\rightarrow\infty$, becomes the well-known Porter-Thomas distribution~\cite{Porter1956}, $P(p) = D_A e^{-D_A p}$, a key signature of the formation of random state ensembles.

\begin{figure}[t!]
	\centering
	\includegraphics[width=89mm]{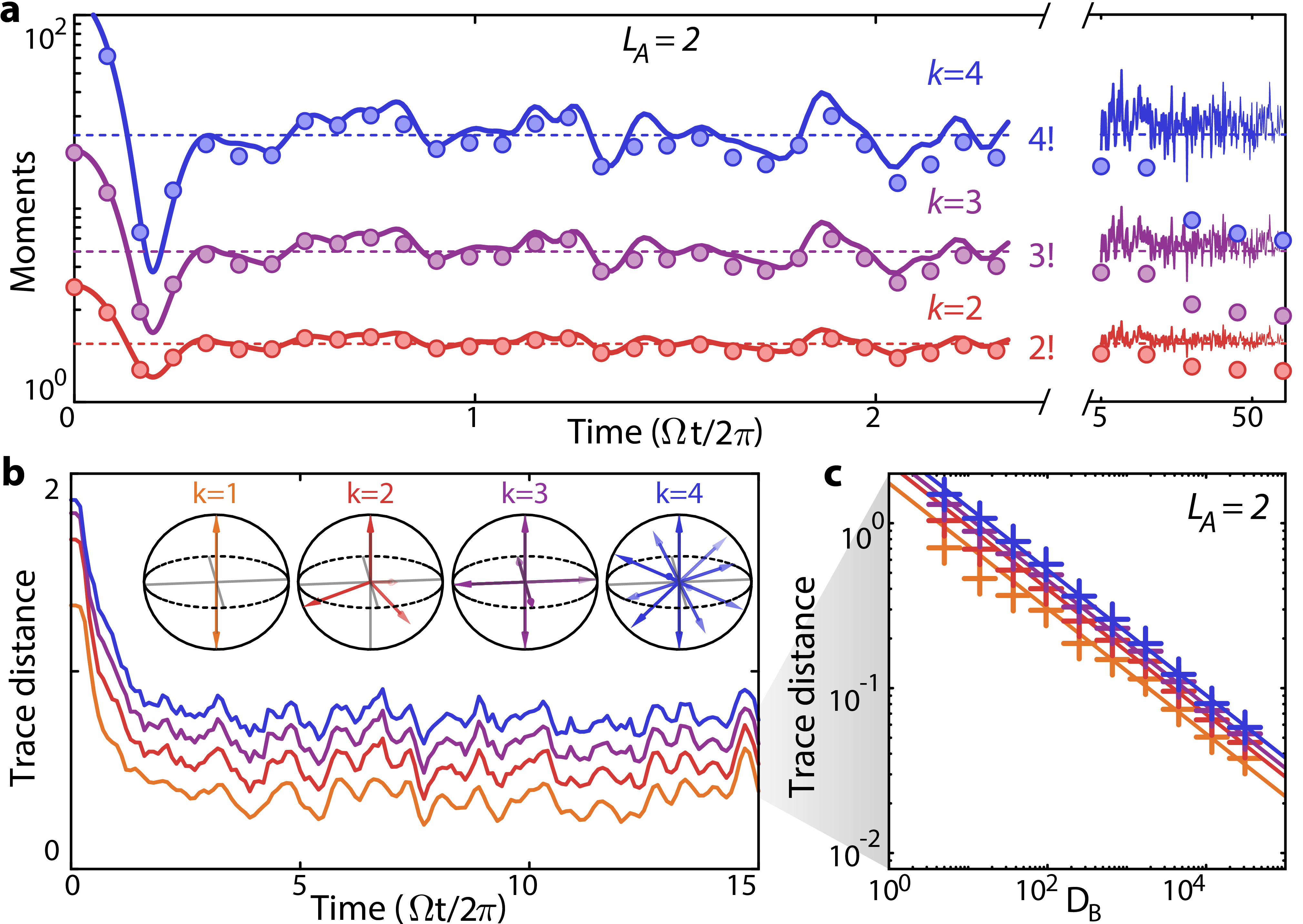}
	\caption{\textbf{Development of emergent randomness. a,} Rescaled second (red), third (purple), and fourth (blue) moments of the conditional probability distributions in Fig.~\ref{Fig2}b for subsystem of length $L_A = 2$. Experimental moments saturate to ${\approx}k!$, the expectation from the uniformly random ensemble (dotted lines) and consistent with numerical simulation (solid lines), before eventually decaying due to decoherence. \textbf{b,} Numerically computed trace distances as a function of time between the $L_A = 2$ projected ensemble and the four lowest order approximations to the uniform random ensemble, so called \textit{quantum state $k$-designs}, for $k=1,2,3,4$ (inset). Distances for all $k$ decrease initially before saturating due to finite system-size effects~\cite{Supplement}. If the trace distances up to order $k$ vanish, the ensemble is as random as the $k\textrm{th}$ design, and fluctuations of observables match up to order $k$, such as the $k\textrm{th}$ moments in a. \textbf{c,} Late-time distances decrease as ${\sim}1/\sqrt{D_B}$ (solid lines), the Hilbert space dimension of the effective bath, subsystem $B$.
\vspace{-0.566cm}} \label{Fig3}
\end{figure}

The convergence of the projected ensemble to a nearly Haar-random distribution can be analyzed in greater detail, and temporarily resolved, by considering moments of the distributions $P(p)$, where the $k$th moment is defined as $p^{(k)}=\sum_p p^k P(p)$ (Fig.~\ref{Fig3}a). Looking order-by-order, we find after rescaling by a factor of $D_A\cdots(D_A+k-1)$, moments from both experiment and numerics quickly approach $k!$, the analytical result expected from a Haar-random ensemble~\cite{Supplement}. Again, at very late time, moments show a characteristic drop, indicating sensitivity to decoherence effects (Fig.~\ref{Fig3}a, right). Crucially, the convergence to $k!$ is independent of the details of subsystem selection, whether $A$ is chosen at the edge, center, or is even discontiguous (Ext. Data Fig.~\ref{EFig_universality}), and universal values are also found for two-point correlators~\cite{Supplement}.  We stress that while the present analysis has been carried out solely for the projected ensemble equilibrated to infinite effective temperature, signatures of similar universal behavior are seen numerically for finite effective temperature cases~\cite{Supplement,Cotler2021}.

Having so far evaluated the projected ensemble solely through the lens of observables, which were consistent with the states being approximately randomly distributed, we now turn to directly quantify the degree of randomness in the projected ensemble by a notion of `distance' not between observables, but between the \textit{ensembles themselves}. To do so, we compare the projected ensemble against progressively more complex approximations to the Haar-random state ensemble, so-called \textit{quantum state k-designs}~\cite{Ambainis2007}. For the case of a single qubit, pictured in the Fig.~\ref{Fig3}b inset, such $k$-designs are increasingly complex distributions of states on the Bloch sphere, realizing the uniform random ensemble for $k\rightarrow\infty$. As our comparison, we take the trace distance between the projected ensemble, generated from error-free simulaton, and successive $k$-designs (Fig.~\ref{Fig3}b); a vanishing distance implies the projected ensemble and the uniform random ensemble are indistinguishable for any observables up to order $k$, including the moments $p^{(k)}$ from Fig.~\ref{Fig3}a. We see that the distances decrease for all $k$th orders as a function of time, before saturating to a value exponentially small in the total system size (Fig.~\ref{Fig3}c). Similar numerical results are found for the case of random unitary circuits and a Hamiltonian used in ion trap experiments (Ext. Data Fig.~\ref{EFig_systems}). In an accompanying paper~\cite{Cotler2021}, we more generally show that the formation of uniformly random, pure state ensembles in subsystems is a more universal phenomenon.

\vspace{0.25cm}
\noindent\textbf{Demonstration of device benchmarking}\newline
A key question is if the formation of approximate $k$-designs in the projected ensemble enables associated applications like device benchmarking with only global, time-independent control. As can be seen in Figs.~\ref{Fig2} and ~\ref{Fig3}, decoherence leads to a noticeable change in observables of the projected ensemble; can this quantitatively determine the onset of decoherence in a quantum device?

We affirmatively answer this question by using the sensitivity of the projected ensemble to decoherence to benchmark the evolution of our experimental system under a time-independent Hamiltonian. Crucially, we stress that our approach would be impossible with access only to the reduced density operator as it is relatively insensitive to decoherence (Fig.~\ref{Fig2}a). As a toy example, we consider the case of a single error occurring at time $t_{\textrm{err}}$ during unitary evolution. The effect of this error then propagates outward~\cite{Khemani2018}, generically transforming the evolution output state and affecting measurement outcomes in subsystem $A$ (Ext. Data Fig.~\ref{EFig_detection}). Using the fact that the projected ensemble forms an approximate $2$-design~\cite{Dankert2009,Boixo2018,Arute2019,Cross2019,Elben2020,Huang2020}, we devise a fidelity estimator $F_c$ to quantify the effect of this error (Methods). The $F_c$ estimator effectively quantifies a rescaled cross-correlation between measurement probabilities in the experimental and ideal conditions:
\begin{align}
F_c=2\frac{\sum_z p_0(z)p(z)}{\sum_z p_0^2(z)}-1,
\label{eq:exactFc}
\end{align}
where $p(z)$ and $p_0(z)$ are the experimental and theoretical probabilities of observing a global bitstring $z$, respectively. We numerically confirm that shortly after we apply an instantaneous phase rotation error on one qubit, our estimator approximates the many-body state overlap, $F_c\approx F=\bra{\psi}\hat{\rho}\ket{\psi}$, between the ideal state, $\ket{\psi}$, and the erroneous state, $\hat{\rho}$ (Ext. Data Fig.~\ref{EFig_detection}b, Methods)~\cite{Supplement}. 

\begin{figure}[t!]
	\centering
	\includegraphics[width=89mm]{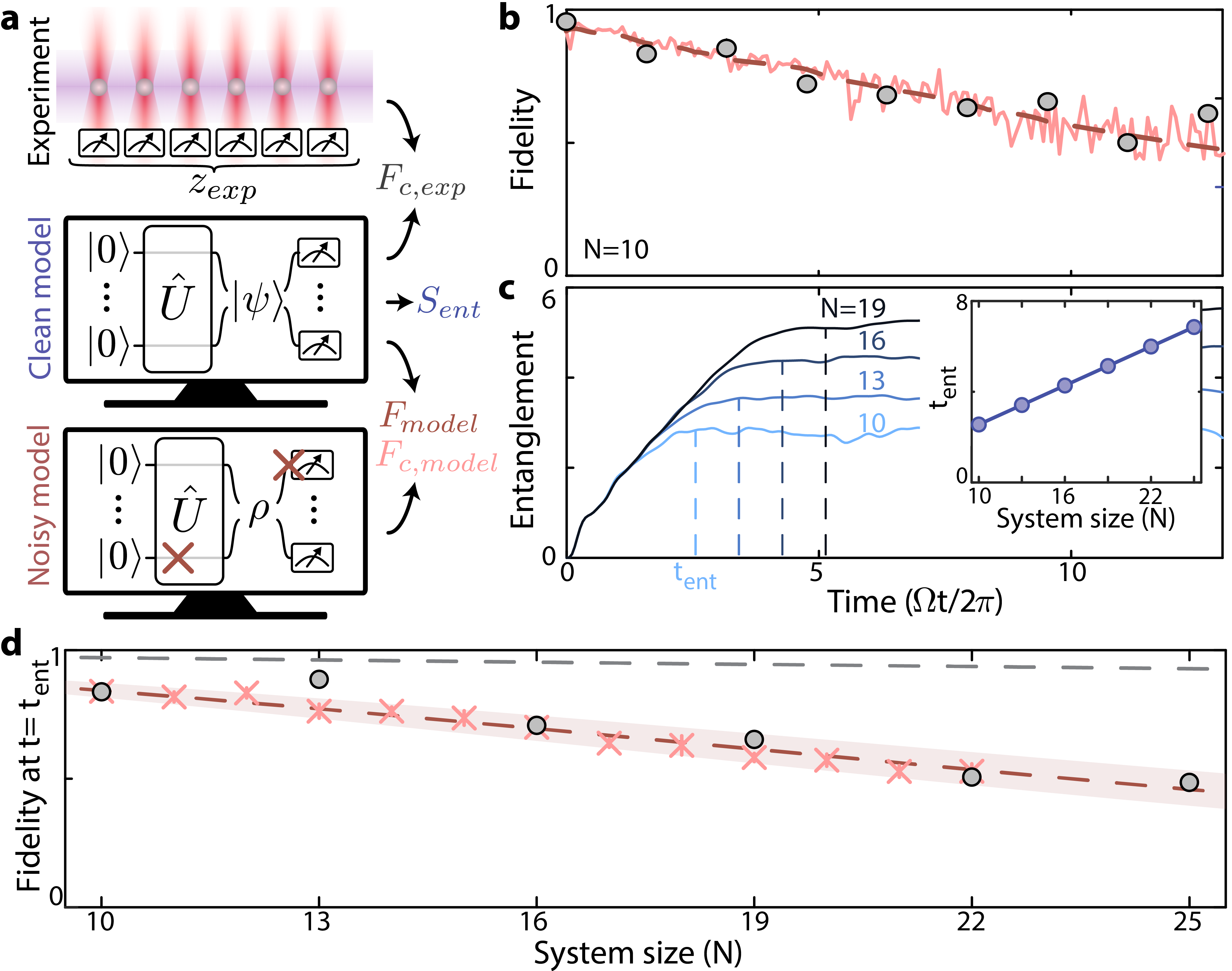}
	\caption{\textbf{Fidelity estimation of an analog Rydberg quantum simulator. a,} To estimate experimental fidelity, we repeatedly perform Hamiltonian evolution, each time performing a projective measurement to accrue an ensemble of measured bitstrings $z_\textrm{exp}$. We then correlate the measured bitstrings with an error-free simulation of the dynamics in order to calculate the fidelity estimator, $F_{c,\textrm{exp}}$. To validate our fidelity estimation method, we compare the error-free simulation against results from an \textit{ab initio} error model~\cite{Supplement}, to calculate the model fidelity $F_\text{model}$ and accompanying estimator $F_{c,\textrm{model}}$. \textbf{b,} Experimental benchmarking of a Rydberg quantum simulator for $N{=}10$ atoms with blockaded Hilbert space dimension $D{=}144$. Shown are $F_{c,\textrm{exp}}$ (grey markers), the fidelity $F_\text{model}$ (dashed red line), and $F_{c,\textrm{model}}$ (solid pink line). \textbf{c,} The half-chain entanglement entropy (calculated from the error-free simulation) increases before saturating at a time, $t_{\mathrm{ent}}$, which grows linearly with system size (inset). \textbf{d,} Fidelity estimated at $t_{\mathrm{ent}}$, showing estimator $F_c$ from experiment (grey markers) up to $N{=}25$, and from error model (pink crosses) up to $N{=}22$. Additionally, we show a fit to the model fidelity, given as $F_0^N \exp[-\gamma(N) t_\text{ent}(N)]$ (red dashed line), where $F_0$ is the single-atom preparation fidelity and $\gamma(N)$ is the many-body fidelity decay rate of our Rydberg simulator (Methods, Ext. Data Fig.~\ref{EFig_fidelitydecay}). The fidelity estimation uses only ${<}10^4$ experimentally sampled bitstrings per data point. See Methods for description of error bars.
	\vspace{-0.566cm}}\label{Fig4}
\end{figure}

To evaluate $F_c$ experimentally, we formulate an empirical, unbiased estimator:
\begin{align}
F_c\approx2\frac{\frac{1}{M}\sum_{i=1}^M p_0(z_\text{exp}^{(i)})}{\sum_z p_0^2(z)}-1,
\label{eq:approxFc}
\end{align}
where $M$ is the number of measurements and $z_\text{exp}^{(i)}$ is the experimentally measured bitstring at the $i$th repetition. While this reformulation still requires calculation of a reference theory comparison, we note that the required number of experimental samples to accurately approximate $F_c$ scales favorably with system size $N$. Concretely, the standard deviation of $F_c$ is estimated to be $\sigma(F_c) \approx 1.04^N/\sqrt{M}$ (Ext. Data Fig.~\ref{EFig_sampling}), meaning that we do not need to fully reconstruct the experimental probability distribution for fidelity estimation of large quantum systems.

We test our benchmarking protocol for errors occurring continuously with a Rydberg quantum simulator of up to $N = 25$ atoms. We estimate the fidelity of our experimental device, $F_{c,\textrm{exp}}$, by correlating measured bitstrings to results from error-free simulation as a function of evolution time. In addition, we use an \textit{ab initio} error model with no free parameters that mimics the experimental output~\cite{Supplement}, from which we extract both the fidelity estimator, $F_{c,\textrm{model}}$, and the model fidelity, $F_\textrm{model}=\bra{\psi(t)}\hat{\rho}_\textrm{model}(t)\ket{\psi(t)}$ (Fig.~\ref{Fig4}a). 

In Fig.~\ref{Fig4}b, we compare $F_\textrm{model}$, $F_{c,\textrm{exp}}$, and $F_{c,\textrm{model}}$ for a system of ten atoms. We observe $F_{c,\textrm{model}}\approx F_\textrm{model}$, validating the efficacy of the estimator under realistic error sources. Additionally, we find $F_{c,\textrm{exp}}\approx F_{c,\textrm{model}}$, and that full bitstring probability distributions show good agreement between the error model and the experiment~\cite{Supplement}, indicating that our \textit{ab initio} error model is a good description of the experiment. 

We further apply this method to estimate the fidelity for generating states with a maximum half-chain entanglement entropy in larger systems. To this end, we first use error-free simulation to calculate the half-chain entanglement entropy growth as a function of system size, finding that the entanglement saturates at a time, $t_{\mathrm{ent}}$, linear in system size (Fig.~\ref{Fig4}c, Methods). We then evaluate the fidelity estimator $F_c$ for $N$ ranging from 10 to 25, each at their respective $t_\mathrm{ent}$, again finding good agreement between experiment and our \textit{ab initio} error model (Ext. Data Fig.~\ref{EFig_fidelitydecay}) in the range for which our error model is readily calculable (Fig.~\ref{Fig4}d). We note an estimated fidelity of 0.49(2) for generating a state with maximum half-chain entanglement entropy for $N=25$.

We numerically show $F_c$ also applies for erroneous evolution using other quantum devices, specifically for random unitary circuits and Hamiltonian evolution in an ion trap quantum simulator (Ext. Data Fig.~\ref{EFig_systems}). In the case of circuits, $F_c$ accurately estimates the fidelity at much shorter evolution times than do existing methods such as linear cross-entropy benchmarking~\cite{Neill2018,Arute2019}, consistent with the early-time formation of the projected ensemble. 

\vspace{0.25cm}
\noindent\textbf{Applications of benchmarking}\newline
Our protocol enables various applications, including evaluating the relative performance of analog and digital quantum devices, \textit{in situ} Hamiltonian parameter estimation, and benchmarking the fidelity of preparing various target states. First, to compare analog and digital quantum evolution, we evaluate the fidelity achieved at $t_{\mathrm{ent}}$ for both analog quantum simulators and digital quantum computers (for which $t_{\mathrm{ent}}$ is defined in terms of gate depth, see Methods). We find our system has an equivalent effective, SPAM-corrected, two-qubit cycle fidelity of 0.987(2) for the gate-set used in Ref.~\cite{Arute2019}, and 0.9965(5) for a gate-set based on two-qubit SU(4) gates~\cite{Cross2019} (Ext. Data Fig.~\ref{EFig_comparison}, Methods).

\begin{figure}[t!!]
	\centering
	\includegraphics[width=89mm]{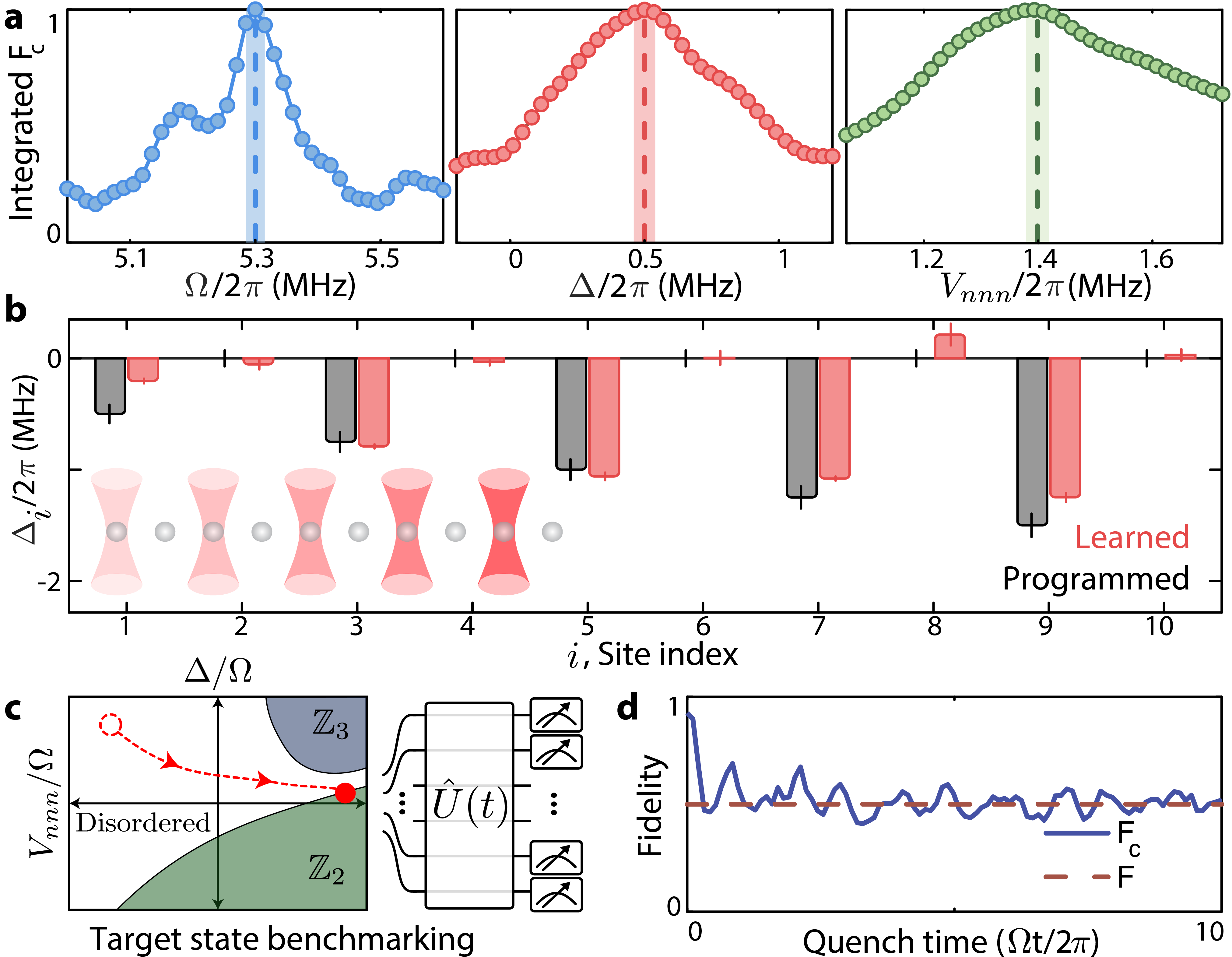}
	\caption{\textbf{Hamiltonian learning and target state benchmarking. a,} Normalized, time-integrated $F_c$ as a function of the global Rabi frequency, detuning, and the next-nearest-neighbor interaction strength in the Rydberg model (Methods); this normalized $F_c$ is maximized only when the correct parameters are used. Vertical dashed lines and shaded areas denote independently calibrated values and their uncertainties. \textbf{b,} Programmed (grey bars) and learned (red bars) local Hamiltonian parameters for an arbitrary, site-dependent detuning field imposed with an intensity-dependent lightshift from locally addressable optical tweezers (inset, red funnels). \textbf{c,d,} Our $F_c$ benchmark (solid line) can estimate the fidelity $F$ (dashed line) of producing a specified target state by evolving at infinite effective temperature after preparation, here numerically demonstrated for a ground state of system size $N=15$ near the $\mathbb{Z}_2$ Ising quantum phase transition in the one-dimensional Rydberg ground state phase diagram~\cite{Slagle2021}, with a noisy state consisting of an equal probability mixture of the ground and first excited states (Methods).
	\vspace{-0.566cm}}\label{Fig5}
\end{figure}

Next, to perform Hamiltonian parameter estimation, we measure $F_c$ while varying Hamiltonian parameters in simulation to find the best agreement between numerical and experimental evolution. For example, we can define a family of target states, which are parameterized by the Rabi frequency, $\Omega$, as \mbox{$\ket{\psi(t,\Omega)}=e^{-i t \hat{H}(\Omega)/\hbar}\ket{0}^{\otimes N}$}. When the value of $\Omega$ does not match the Rabi frequency used in the experiment, the target state $\ket{\psi(t,\Omega)}$ will have smaller overlap with the experimental state, and the fidelity estimator \mbox{$F_c(t,\Omega)\approx \bra{\psi(t,\Omega)}\hat{\rho}(t)\ket{\psi(t,\Omega)}$} will decay more quickly. To capture this effect in a single quantity we plot the normalized, time-integrated $F_c$ (Fig.~\ref{Fig5}a). For each Hamiltonian parameter, a sharp maximum emerges~\cite{Supplement}, showing good agreement with precalibrated values (dashed lines and shaded areas). Parameter estimation also works when applied to learn local, site-dependent terms of a disordered Hamiltonian (Fig.~\ref{Fig5}b), notably without any local control during readout. 

Finally, $F_c$ can be used to benchmark the fidelity of preparing various quantum states of interest by preparing a target state and then quenching the Hamiltonian to evolve the prepared state at infinite effective temperature (Fig.~\ref{Fig5}c, Methods). As a numerical proof-of-principle, we show results for such target state benchmarking to prepare a ground state near the Ising quantum phase transition in the Rydberg model~\cite{Fendley2004,Slagle2021} (Fig.~\ref{Fig5}c,d), where the noisy state is an equal probability mixture of the ground and first excited states. After a short disordered quench, the estimator $F_c$ reveals the fidelity of the prepared state, offering a novel way to perform \textit{in situ} optimization of many-body state preparation; further examples are shown in Ext. Data Fig.~\ref{EFig_tsb}.

In conclusion, we have uncovered emergent randomness arising from partial measurement of an interacting many-body system, and have subsequently shown a widely applicable fidelity estimation scheme which works at shorter evolution times and with reduced experimental complexity compared to existing approaches. We have further demonstrated applications in quantum device comparison, Hamiltonian parameter estimation, and benchmarking the fidelity of preparing interesting quantum states. The concept of emergent randomness could provide a new framework for quantum thermalization, chaos, and complexity growth~\cite{Cotler2020}. Open questions remain, such as a deeper understanding of the finite effective temperature case~\cite{Cotler2021,Supplement}, and uncovering the signatures of non-ergodic dynamics in integrable or localized systems~\cite{Rigol2008,Nandkishore2015,Abanin2019,Ueda2020,Turner2018}. Such developments could enable a more flexible and standardized way of performing quantum fidelity estimation in a wide variety of quantum hardware, including trapped ions~\cite{Monroe2019}, superconducting qubits~\cite{Neill2018}, photonic systems~\cite{Zhong2020}, solid-state spins~\cite{Zwanenburg2013,Awschalom2018}, and cold atoms and molecules in optical lattices~\cite{Gross2017}.  Ultimately, emergent random ensembles could find a broader range of applications, including quantum advantage tests~\cite{Boixo2018,Arute2019,Bouland2019,Haferkamp2020,Zhong2020,Wu2021}, \textit{in situ} Hamiltonian learning~\cite{Giovannetti2004,Arute2019}, and optimization of target quantum state preparation.
 
\textit{Note added} -- During the course of the revision, a new fidelity estimator has been introduced~\cite{Mark2022}; we present a comparison in Ref.\cite{Supplement}.

\FloatBarrier
\bibliographystyle{manubib2}
\bibliography{library_endreslab.bib}

\FloatBarrier

\newpage

\setcounter{figure}{0}
\captionsetup[figure]{labelfont={bf},name={Ext. Data Fig.},labelsep=bar,justification=raggedright,font=small}
\clearpage
\section*{Extended Data Figures}
\FloatBarrier

\FloatBarrier

\begin{figure}[b!]
	\centering
	\includegraphics[width=70.3mm]{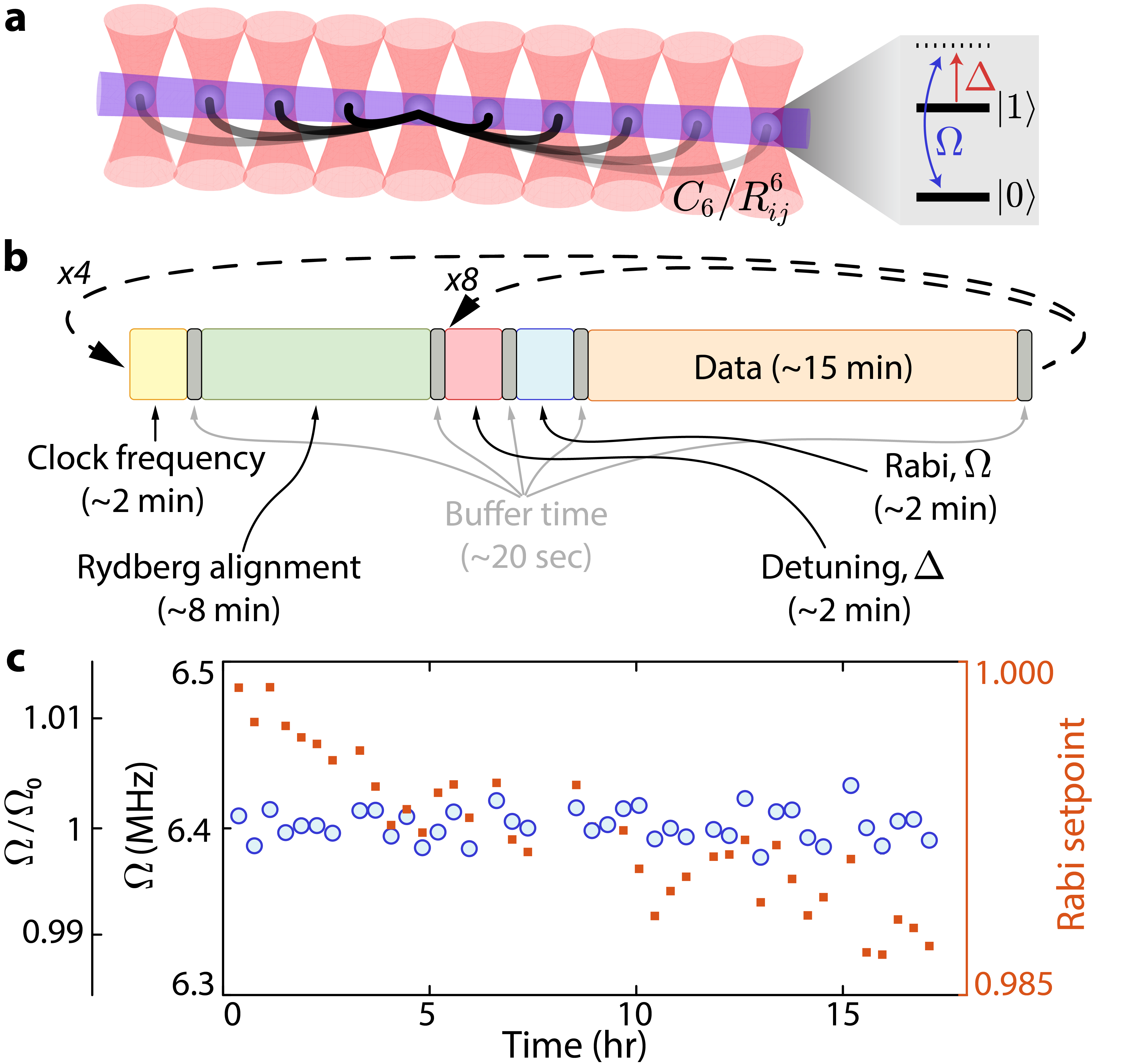}
	\caption{\textbf{Experimental system and parameter feedback:} \textbf{a,} Illustration of a Rydberg quantum simulator consisting of strontium-88 atoms trapped in optical tweezers (red funnels). All atoms are driven by a global transverse control field (purple horizontal beam) at a Rabi frequency $\Omega$ and a detuning $\Delta$ (right panel). The interaction strength is given as $C_6/R_{ij}^6$ with an interaction constant $C_6$ and atomic separations $R_{ij}$ between two atoms at site $i$ and $j$. \textbf{b,} Schematic of the experimental feedback scheme. We automatically interleave data taking with feedback to global control parameters and systematic variables through a home-built control architecture (Methods); in particular, we feedback to the clock laser frequency (to maintain optimal state preparation fidelity), the Rydberg laser alignment, the Rydberg detuning $\Delta$, and the Rabi frequency $\Omega$. \textbf{c,} Example of the interleaved automatic Rabi frequency stabilization over the course of ${\approx} 20$ hours with no human intervention. Feedback is comprised of performing single-atom Rabi oscillations, fitting the observed Rabi frequency, and updating the laser amplitude, rather than simply stabilizing the laser amplitude against a photodiode reference. While the Rabi frequency setpoint (orange squares) changes over the course of the sequence (due to long-time instabilities like temperature drifts), the measured Rabi frequency (blue circles) stays constant to within ${<}0.3\%$, with a standard deviation of $0.15\%$. This same stability is seen over the course of multiple days with nearly continuous experimental uptime.
	} 
	\label{EFig_feedback}
\end{figure} 

\begin{figure}[t!]
	\centering
	\includegraphics[width=89mm]{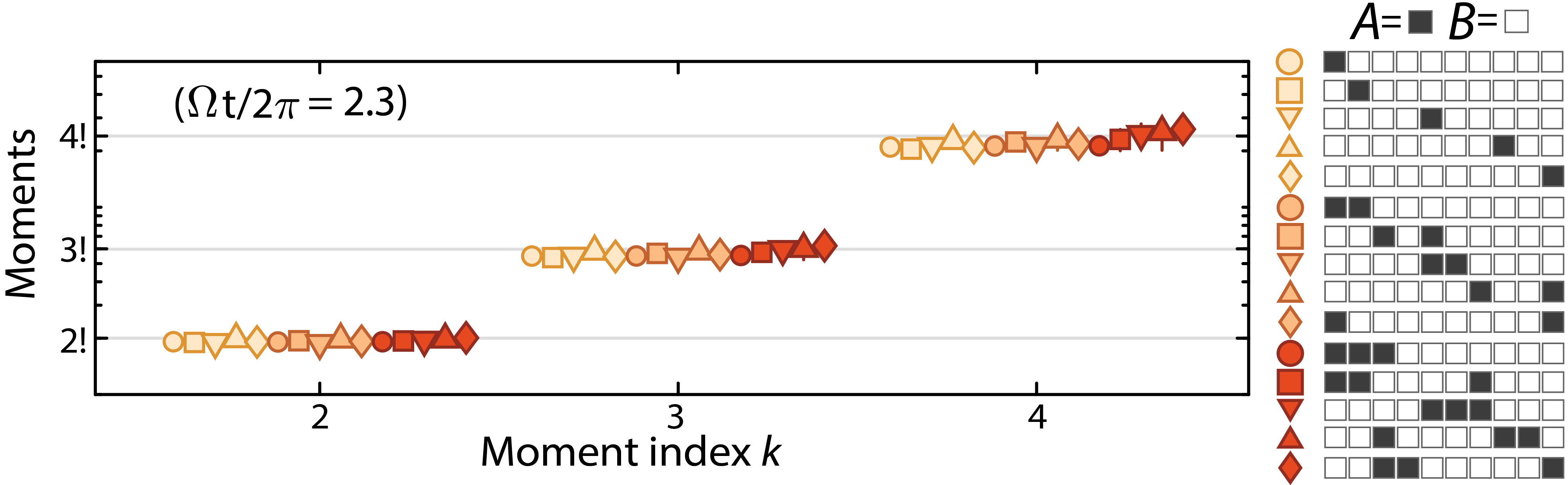}
	\caption{\textbf{Universality of moments of the projected ensemble.} $k\textrm{th}$ moments of the conditional probability distributions in Fig.~\ref{Fig2}b,c, evaluated at late-time ($\Omega t/2\pi=2.3$) and for a variety of choices of subsystems (see panel on the right); we find a universal convergence to ${\approx}k!$, independent of subsystem choice, suggesting that a subsystem's projected ensemble converges to the uniform random ensemble irrespective of the details of placement, or connectivity. Error bars are the standard deviation over temporal fluctuations in moments near the evaluated time, as shown in Fig.~\ref{Fig3}a.} 
	\label{EFig_universality}
\end{figure} 

\begin{figure}[t!]
	\centering
	\includegraphics[width=89mm]{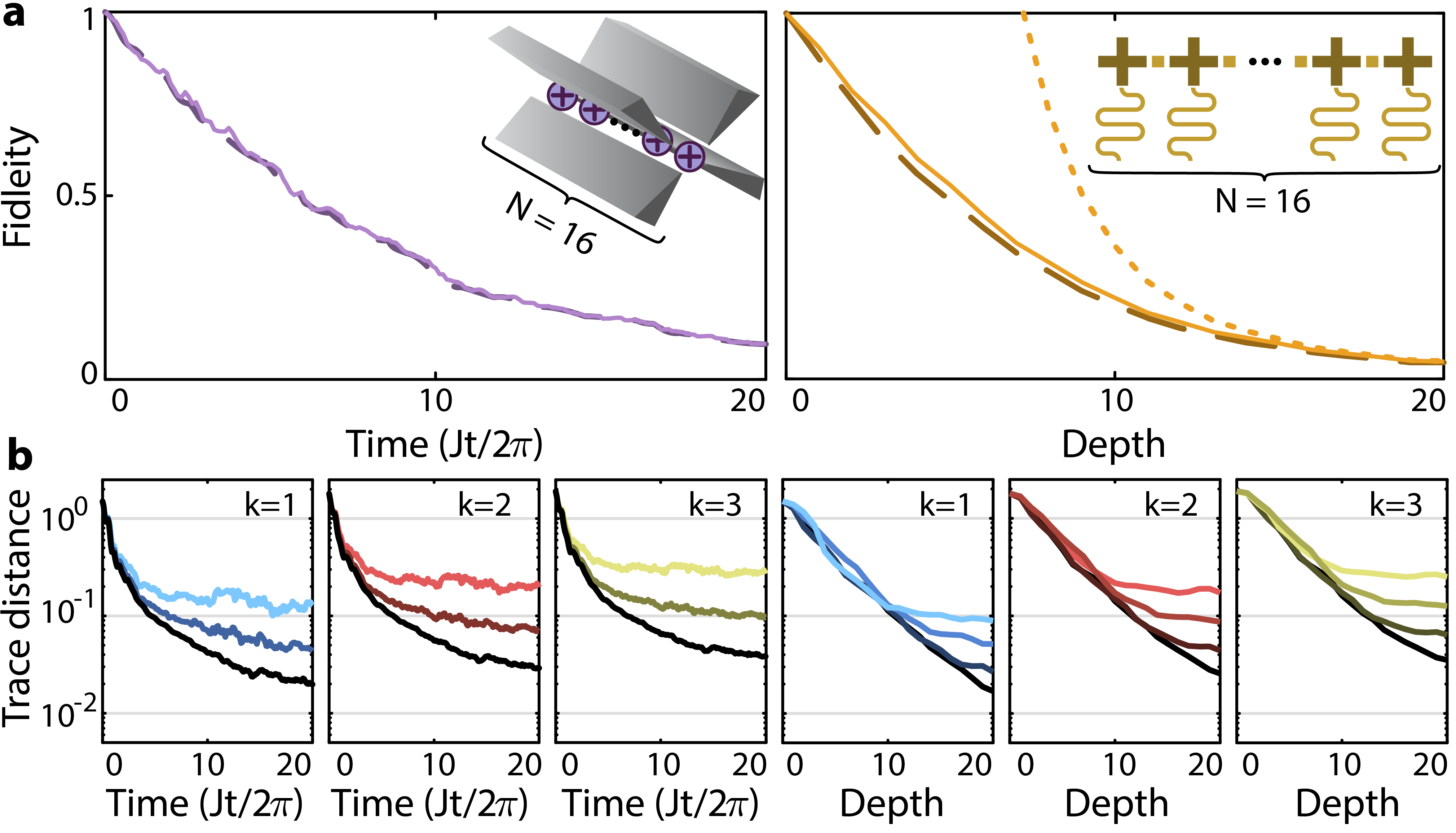}
	\caption{\textbf{Emergent randomness and benchmarking in other quantum systems. a,} Fidelity estimation for the case of a trapped ion quantum simulator governed by chaotic Hamiltonian evolution (left) and a quantum computer implementing a random unitary circuit (RUC) (right); see Ref. ~\cite{Supplement} for simulation details. In both cases, we plot the many-body fidelity (dashed line), as well as our fidelity estimator, $F_c$ (solid line); for the RUC case we also plot the more conventional linear cross-entropy-benchmark, $F_\textrm{XEB}$~\cite{Arute2019} (dotted line). We find that $F_c$ approximates the fidelity at much earlier times than $F_\textrm{XEB}$. \textbf{b,} Numerically computed trace distances between the projected ensemble of a two-qubit subsystem and the corresponding $k$-design. Results are shown for multiple different total system sizes: 10, 13, 16 for the trapped ion case, and 10, 12, 14, 16 for the RUC case, with darker colors corresponding to larger total system sizes.} 
	\label{EFig_systems}
\end{figure} 

\begin{figure}[t!]
	\centering
	\includegraphics[width=89mm]{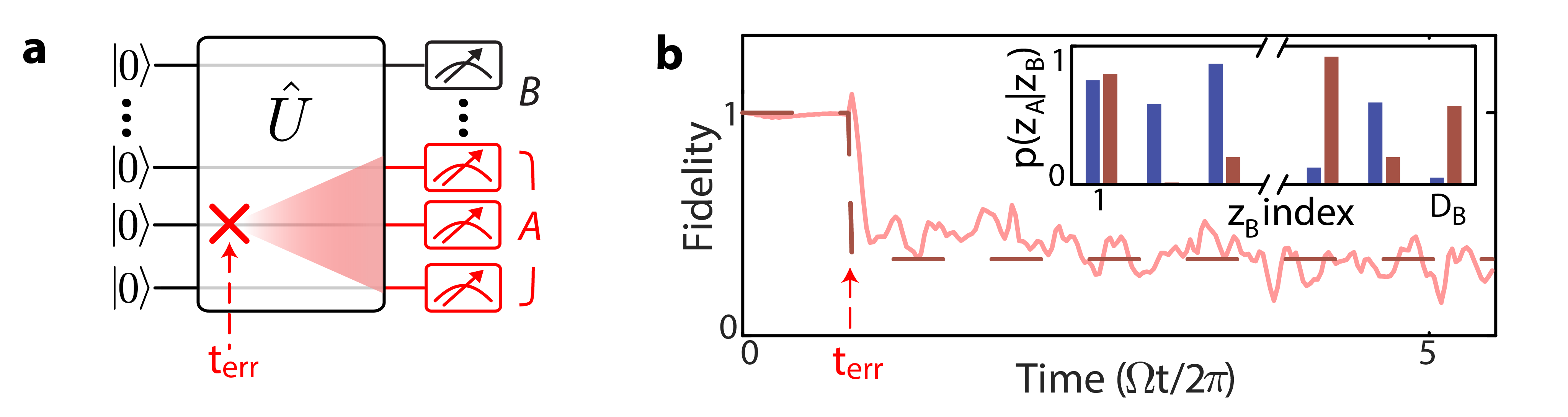}
	\caption{\textbf{Detecting errors during quantum evolution. a,} Schematic of noisy time evolution with an error occurring at time $t_\mathrm{err}$. The influence of the local error propagates outward, affecting the measurement outcomes non-locally at a later time. \textbf{b,} Errors during evolution can be detected by correlating the measurement outcomes with an error-free, ideal evolution case. We numerically tested this by applying a local, instantaneous phase error to the middle qubit of an $N=16$ atom Rydberg simulator at time $\Omega t_\mathrm{err}/2\pi \approx 1$. The proposed fidelity estimator, $F_c$ (solid line), accurately approximates the many-body overlap (dashed line) between states produced with and without errors, after a slightly delayed time. Inset: Conditional probability distributions in $A$ before (blue) and after (red) the error, showing decorrelation.
	}\label{EFig_detection}
\end{figure}

\begin{figure}[t!]
	\centering
	\includegraphics[width=89mm]{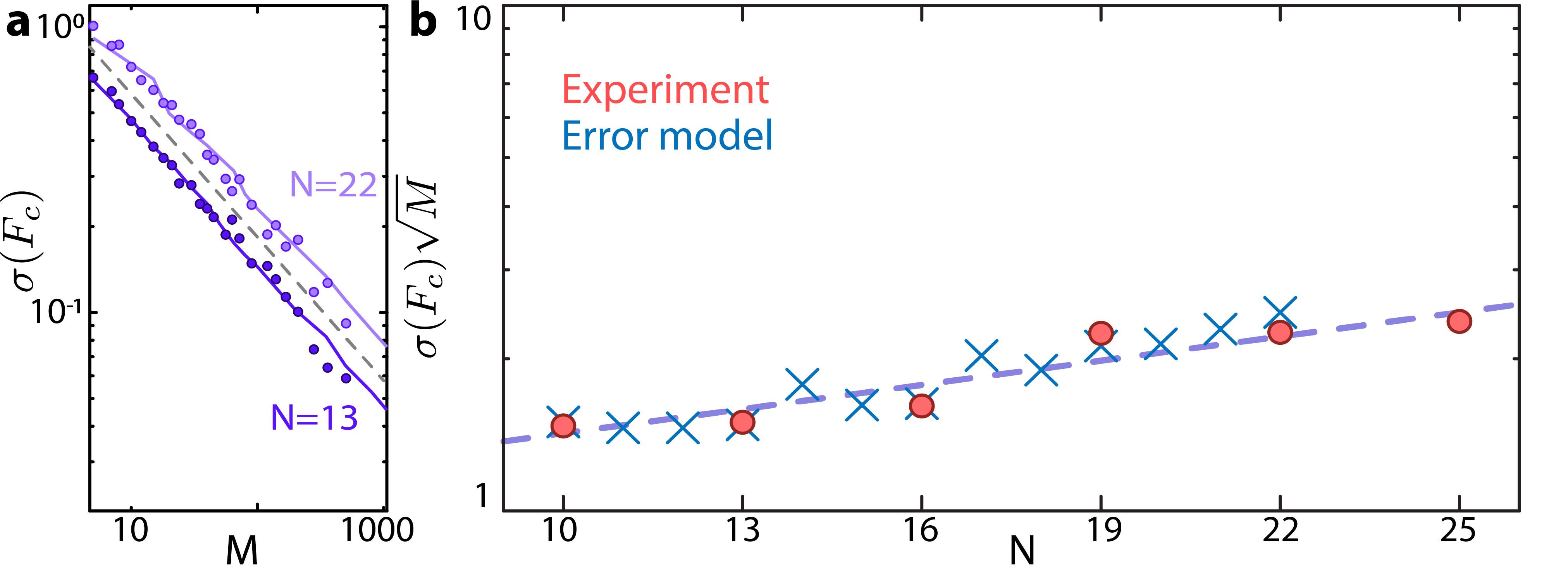}
	\caption{\textbf{Finite sampling analysis for $F_c$. a,} Statistical fluctuations of the fidelity estimator, $F_c$, at $N{=}13$ (dark purple) and $N{=}22$ (light purple), computed both using our \textit{ab initio} error model (solid lines) and experiment (markers) evaluated with a finite number of $M$ bitstring samples. Data are consistent with a $1/\sqrt{M}$ scaling, shown here as a guide to the eye (grey dashed line) \textbf{b,} Sample complexity of the fidelity estimator, evaluated at the $N$-dependent entanglement saturation time for the error model (blue crosses), and for the experimental data in Fig.~\ref{Fig4}d (red circles). A fit to the experimental data (dashed line) with functional form $\sigma(F_c)\sqrt{M}=a^N$ yields an estimate of $a=1.037(2)$ (a similar fit to the error model yields an estimate of $a=1.039(2)$).} 
	\label{EFig_sampling}
\end{figure} 

\begin{figure}[t!]
	\centering
	\includegraphics[width=89mm]{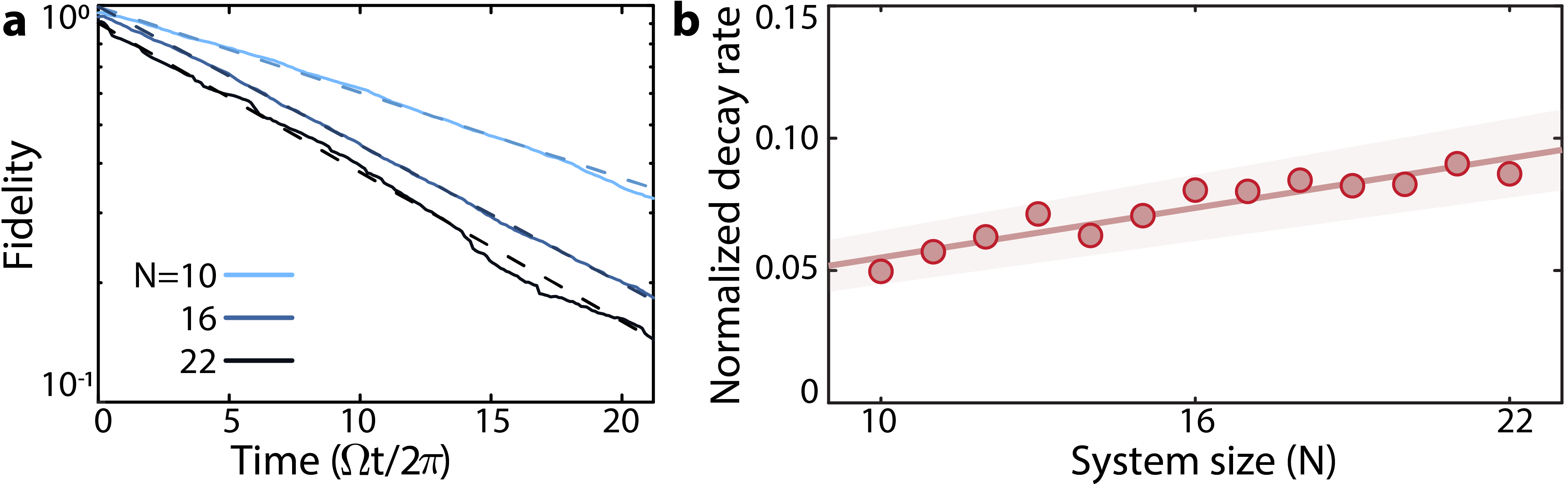}
	\caption{\textbf{Predicting fidelity scaling. a,} We use our \textit{ab initio} error model (which includes state preparation errors) to predict the fidelity decay rate as a function of system size. For various system sizes we plot the model fidelity (solid lines), as well as fits to exponential decay with an unconstrained value at $t=0$ (dashed lines), which we see are consistent with the time-dependent fidelity. \textbf{b,} For the range of system sizes for which our error model is readily calculable, we see the fidelity decay rate, $\gamma(N)$ (markers), is consistent with a linear function of system size (red line). The shaded region comes from uncertainty in the fit parameters.
	}\label{EFig_fidelitydecay}
\end{figure}

\begin{figure*}[t!]
	\centering
	\includegraphics[width=125mm]{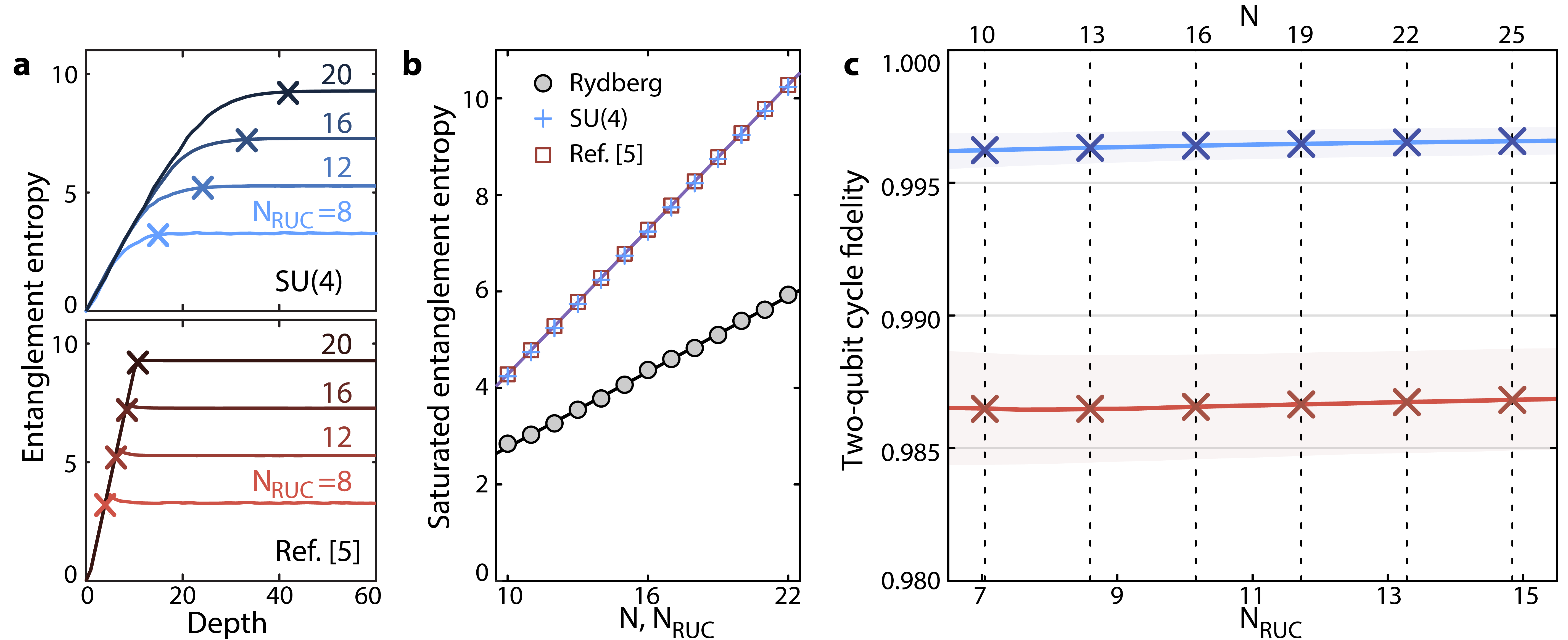}
	\caption{\textbf{Comparison to digital quantum devices executing random circuits. a,} Numerical simulations of a one-dimensional digital quantum device implementing a random unitary circuit (RUC). Two different digital gate implementations are tested: a configuration based on the gate-set used in Ref.~\cite{Arute2019} (bottom), and an configuration where each cycle is composed of parallel two-qubit SU(4) gates (top)~\cite{Cross2019}. Cross markers indicate when the half-chain entanglement entropy saturates. \textbf{b,} Due to the Rydberg blockade mechanism, as well as symmetries of the Rydberg Hamiltonian~\cite{Supplement}, an equal number of atoms in the Rydberg simulator, $N$, and qubits in the RUC, $N_\mathrm{RUC}$, will not saturate to the same half-chain entanglement entropy. However, we can still find an equivalence by plotting the saturated entanglement entropy for the RUC (blue crosses for the SU(4) gate-set, open red squares for gate-set from Ref.~\cite{Arute2019}) and for the Rydberg simulator (grey markers) as a function of their respective system sizes. We fit the results for the Rydberg simulator (black line), and plot the analytic prediction for the RUC~\cite{Page1993} (purple line), from which we can write an equivalent $N_\mathrm{RUC}$ as a function of $N$, in the sense of maximum achievable entanglement entropy (Methods). \textbf{c,} For a given $N$ (and equivalent $N_\mathrm{RUC}$), we plot the SPAM-corrected, two-qubit cycle fidelity for an equivalently-sized RUC to match the evolution fidelity of our Rydberg simulator at the time/depth when entanglement saturates. Red lines, markers and crosses are for the gate-set of Ref.~\cite{Arute2019}, while blue are for the SU(4) gate-set. Shaded regions come from the error on fitting the various $N$-dependent parameters which enter this calculation (Methods).}
	\label{EFig_comparison}
\end{figure*} 

\begin{figure}[b!]
	\centering
	\includegraphics[width=89mm]{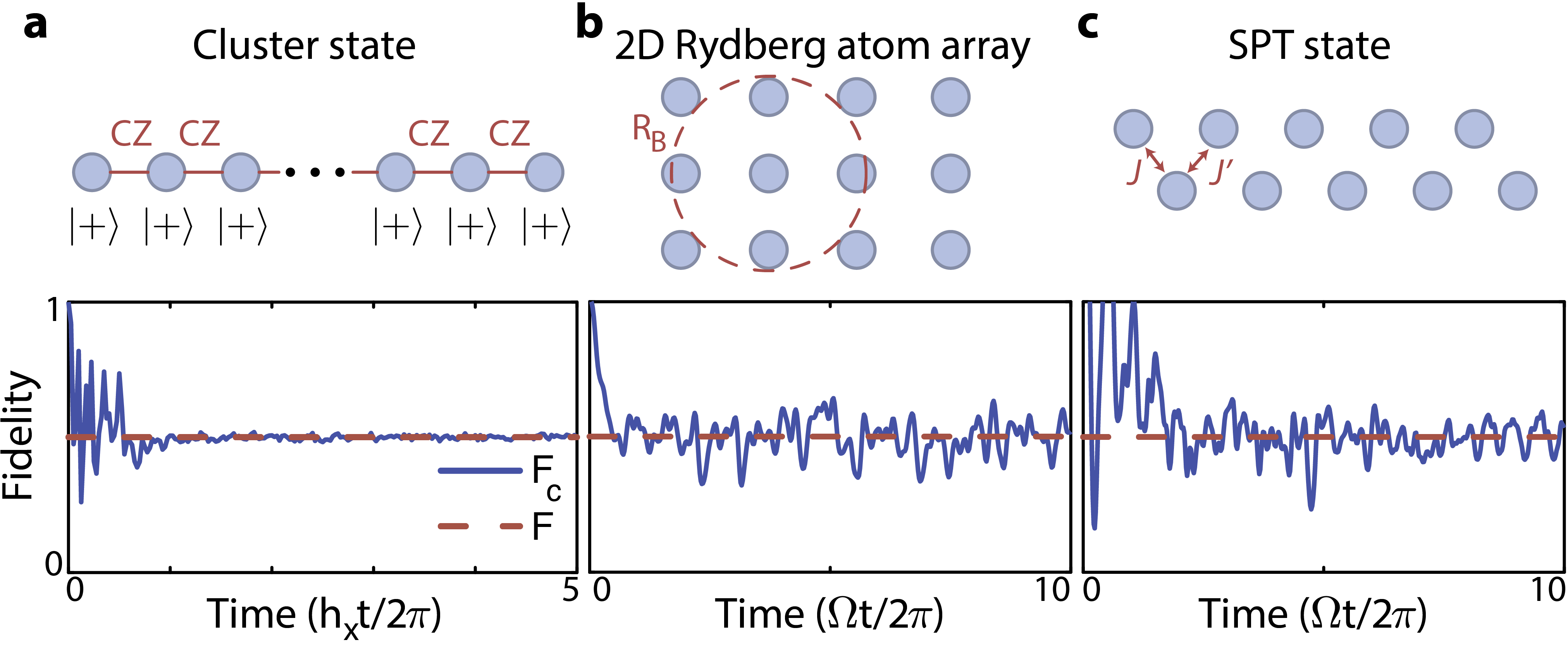}
	\caption{\textbf{Applications to target state benchmarking. a,} Benchmarking of a one-dimensional cluster state, \textbf{b,} a pure Haar-random state benchmarked in a two-dimensional square Rydberg atom array, and \textbf{c,} a symmetry-protected topological (SPT) ground state prepared in a Rydberg ladder array realizing the Su-Schrieffer-Heeger topological model~\cite{de2019}. In \textbf{a}, CZ denotes a controlled-$Z$ gate and $\ket{+} = \frac{\ket{0}+\ket{1}}{\sqrt{2}}$. In \textbf{b}, $R_B$ denotes the Rydberg blockade radius within which more than a single excitation is not allowed~\cite{Bernien2017,Browaeys2020,Madjarov2020}. In \textbf{c}, $J$ and $J'$ are the alternating coupling strengths of a two-leg ladder array. In all cases, $N=16$ qubits are used, and imperfect quantum states are prepared via phase rotations such that the many-body fidelity overlap becomes 0.5 (red dashed line). Additionally, chaotic evolution is performed such that the initial state is at infinite effective temperature to apply our $F_c$ formalism (blue solid lines) (Methods).} \vspace{-0.5cm}
	\label{EFig_tsb}
\end{figure}

\FloatBarrier

\section*{Methods}

\subsection*{Description of the Experiment}
The details of our experiment have been summarized previously\cite{Cooper2018,Covey2019b,Madjarov2019,Madjarov2020}; in brief, we use an array of optical tweezers to trap individual strontium-88 atoms. Initially in the $5s^2$ $^{1}S_{0}$ state, atoms are cooled on the narrow-line $5s^2$ $^{1}S_{0}$ $\leftrightarrow$ $5s5p$ $^{3}P_{1}$ (689~nm) transition close to their motional ground state, with an average transverse occupation number of $\langle n\rangle{\approx}0.3$ (corresponding to ${\approx}$3 $\mu$K). For all data shown, we rearrange the initially stochastically filled array to a defect-free array~\cite{Barredo2016,Endres2016} of atoms spaced by 3.75 $\mu$m, discarding extras. Atoms are initialized to the $5s5p$ $^{3}P_{0}$ (698~nm) \textit{clock state} through a combination of coherent drive and incoherent pumping, for a total preparation fidelity of 0.997(1) per atom. We treat the clock state as a metastable qubit ground state, $|0\rangle$, and subsequently drive to the $5s61s$ $^{3}S_{1},m_{J}{=}0$ (317~nm) Rydberg state, $|1\rangle$. Following Hamiltonian evolution, state readout is performed using the auto-ionizing transition $5s61s$ $^{3}S_{1},m_{J}{=}0$ $\leftrightarrow$ $5p_{3/2}61s_{1/2}$ (408~nm, $J{=}1,m_{J}{=\pm}1)$ which rapidly ionizes atoms in the Rydberg state with high fidelity (${\approx}0.999$), leaving them dark to our fluorescent imaging. Atoms in the clock state are pumped into the imaging cycle, allowing us to directly map atomic fluorescence to qubit state.

The Hamiltonian of this system is well approximated by
\begin{align}
\hat{H}/\hbar=\Omega\sum_i \hat{S}_i^x -\Delta\sum_i \hat{n}_i + \frac{C_6}{a^6} \sum_{i>j} \frac{\hat{n}_i \hat{n}_j}{|i-j|^6} 
\end{align}
which describes a set of interacting two-level systems, labeled by site indices $i$ and $j$, driven by a laser with Rabi frequency $\Omega$ and detuning $\Delta$. The interaction strength is determined by the $C_6$ coefficient and the lattice spacing $a$. Operators are $\hat{S}^x_i =(\ket{1}_i\bra{0}_i + \ket{0}_i \bra{1}_i)/2$ and $\hat{n}_i = \ket{1}_i\bra{1}_i$, where $\ket{0}_i$ and $\ket{1}_i$ denote the electronic ground and Rydberg states at site $i$, respectively. 

For measurements observing the emergence of random ensembles, we use $\Omega/2\pi = 4.70(1)$ MHz, $\Delta/2\pi = 0.90(3)$ MHz, $a=3.75(5)$ $\mu$m, with an experimentally measured next-nearest-interaction strength of $V_\text{nnn}/2\pi=C_6/(2a)^6=1.40(2)$ MHz, yielding an estimated $C_6$ coefficient of $2\pi\times249(20)$ GHz~$\mu$m$^6$. Under this condition, we confirm numerically that the initial all-zero state rapidly thermalizes to an infinite effective temperature thermal ensemble locally within the constrained subspace where no two adjacent atoms are simultaneously in the Rydberg state~\cite{Bernien2017,Browaeys2020,Madjarov2020}. Benchmarking measurements are performed with $\Omega/2\pi = 5.30(1)$ MHz, $\Delta/2\pi = 0.50(3)$ MHz, which again thermalizes to an infinite effective temperature thermal ensemble.

As the experimental data shown throughout the main text requires both high statistics (taken over the course of multiple days) and very fine parameter control, we periodically perform automatic feedback to several experimental parameters using a home-built control architecture. Specifically these are: 1) the clock state resonance frequency to ensure maximal preparation fidelity, 2) the Rydberg laser beam alignment, 3) the Rydberg resonance frequency, and 4) the Rydberg Rabi frequency. For the clock frequency, we apply a $\pi$-pulse on the clock transition to identify the resonance and perform state-resolved readout by ejecting all ground state atoms from the trap with an intense pulse of light on the $5s^2$ $^{1}S_{0}$ $\leftrightarrow$ $5s5p$ $^{1}P_{1}$ (461~nm) transition~\cite{Madjarov2020}. 

For the Rydberg alignment, detuning, and Rabi frequency, we rearrange the array to \textit{non-interacting} atoms spaced by $15.1~\mu$m. During alignment we raster the Rydberg beam across the array sampling different position-dependent Rabi frequencies, and thus evolving to different position-dependent phases. We compare the resultant signal across all positions to a simulation to identify the point of furthest phase, and thus maximal intensity. For the Rydberg detuning, we measure the resonance condition at $\Omega t=13\pi$ in order to narrow the resonance feature. For the Rabi frequency, we take a series of time points between $13\pi < \Omega t < 17\pi$, and fit the resulting Rabi oscillations. After each feedback experiment, the relevant parameter is automatically updated for subsequent measurements (Ext. Data Fig.~\ref{EFig_feedback}).

\subsection*{Data Analysis}
\label{sec:dataanalysis}
Our state readout is described in detail in Ref.~\cite{Madjarov2020}; it features single-site detection which discriminates atoms in the clock state, $|0\rangle$, versus the Rydberg state, $|1\rangle$, through a combination of fluorescence imaging and Rydberg auto-ionization. We take a total of three images: 1) after the array is initially loaded to perform rearrangement, 2) after the rearrangement is completed to verify the initial state is correct, 3) after the sequence has finished. We post-select for image triplets where the proper rearrangement pattern is visible in image (2), and calculate the survival of each atom by comparing site occupations in image (2) to image (3). This array of survival signals is then converted into the qubit basis. For instance, in typical experiments where atoms are rearranged into defect-free arrays of ten atoms, we calculate the binary survivals for each atom, and then make the mapping `atom survived'$\rightarrow |0\rangle$ and `atom did not survive'$\rightarrow |1\rangle$, yielding a bitstring of the qubit states. After taking many shots we accrue an ensemble of such bitstrings, $\{z\}$. For randomness measurements, a total of ${\approx}120000$ shots are used (${\approx}3000$ shots per time point). For benchmarking measurements a total of ${\approx}40000$ shots are used for generating the time-trace at $N=10$ in Fig.~\ref{Fig4}b (${\approx}3700$ shots per time point). Approximately ${\approx}44000$ total shots are used for the $N$-scaling plot in Fig.~\ref{Fig4}d, where the number of shots for a given system size is approximately given by $M\approx3000+250N$.

Error bars in Figs.~\ref{Fig2}, and ~\ref{Fig3}, are calculated via bootstrapping methods, and are often smaller than the marker sizes. In Fig.~\ref{Fig4}, error bars on experimental quantities are calculated via extrapolation from subsampling of the total number of experimentally measured bitstrings to estimate the sample complexity at a given $N$ (Ext. Data Fig.~\ref{EFig_sampling}). The error bars on $F_{c,\text{model}}$ from the {\it ab initio} error model stem from typicality errors associated with the temporal fluctuation of our estimator~\cite{Supplement}. Error bars on the programmed parameters in Fig.~\ref{Fig5}b come from uncertainty in local detuning intensity, while error bars on the learned parameters are standard deviations arising from performing the simultaneous parameter optimization 30 times with randomized starting initial conditions.

Our system Hamiltonian is naturally stratified into a number of energetically widely spaced sectors due to the Rydberg blockade~\cite{Bernien2017,Browaeys2020,Madjarov2020}. In particular, the nearest-neighbor interaction is ${\approx}20\times$ greater than the next largest energy scale, so cases where neighboring pairs of atoms are both excited to the Rydberg state are greatly suppressed. For $N=10$, we find ${\approx}99\%$ of all experimental bitstrings are in the blockade-satisfying energy sector at short times ($t < 1$ $\mu$s) but this probability starts to decrease at late times ($t > 1$ $\mu$s) due to experimental imperfections - we refine $\{z\}$ by discarding all realizations not in this sector. We note, however, that all simulations are performed in the full Hilbert space.

For calculations involving conditional probabilities, we bipartition each bitstring $z$ into subsystems $A$ and $B$ with bitstrings $z_A$ and $z_B$ respectively. When considering the statistics of conditional probabilities, we note that the blockade interaction can reduce the dimensionality of the Hilbert space of subsystem $A$ if the boundary qubits in $B$ are in the Rydberg state. To isolate a set of conditional states having the same Hilbert space dimension, $D_A$, for a given choice of subsystem $A$ and $B$, we only consider bitstrings $z_A$ and $z_B$ if the qubits in $B$ bordering $A$ are in the $|0\rangle$ state. 

\subsection*{Derivation of the fidelity estimator $F_c$}
\label{sec:fc_derivation}
Our fidelity estimator $F_c$ (Eq.~\ref{eq:exactFc}) can be understood by expressing the global bitstring probabilities for ideal and noisy evolutions, $p_0(z)$ and $p(z)$ respectively, in terms of conditional and marginal probabilities as
\begin{align}
 p_0(z) &= p_0(z_A|z_B) p_0(z_B) \\ 
 p(z) &= p(z_A|z_B) p(z_B),
 \end{align}
for complementary subsystems $A$ and $B$. We consider the simplest case of a single local error $\hat{V}$ occurring at time $t_\textrm{err}$ during time-evolution, and assume that the time-evolved error operator, $\hat{V}(\tau)=\hat{U}(\tau)\hat{V}\hat{U}(\tau)^\dagger$, is supported within subsystem A. Here $\tau=t-t_\textrm{err}$ is the time past the occurrence of the error and $\hat{U}(\tau)$ is the time-evolution operator from $t_\textrm{err}$ to $t$. This implies that the measurement outcome in $B$ is not affected by the error, giving $p(z) = p(z_A|z_B) p_0(z_B)$ because $p(z_B) = p_0(z_B)$. Under these conditions, we can rewrite $F_c$ as 
\begin{align}
F_c &= 2 \frac{\sum_z p_0(z) p(z)}{\sum_z p_0(z)^2} - 1 \\
&=2 \frac{\sum_{z_B} p_0^2 (z_B) \sum_{z_A} p(z_A|z_B) p_0 (z_A|z_B)}{\sum_{z_B} p_0^2 (z_B) \sum_{z_A} p_0^2(z_A|z_B)} - 1\\
\label{method_eqn:avg_XEB}
&\approx \frac{\sum_{z_B} p_0^2 (z_B)  (F_\text{XEB} (z_B) + 1) }{\sum_{z_B} p_0^2 (z_B)} - 1  \\
&= \sum_{z_B} q(z_B) F_\text{XEB} (z_B) \label{eq:Fc_deriv_full}
\end{align}
where $q(z_B) = \frac{p_0^2(z_B)}{\sum_{z_B} p_0^2(z_B)}$ and
\begin{align}
F_\text{XEB}(z_B) = (D_A+1) \sum_{z_A} p(z_A|z_B) p_0 (z_A|z_B) -1
\end{align}
is the $z_B$-dependent, linear cross-entropy benchmark~\cite{Arute2019} in subsystem $A$, and $D_A$ is the Hilbert space dimension of $A$. From Eq.~(8) to Eq.~(9), we used the second-order moment of the projected ensemble in an error-free case 
\begin{align}
	\frac{1}{D_A} \sum_{z_A} p_0^2(z_A |z_B) \approx \frac{2!}{D_A(D_A+1)}
	\label{eq:p2_equiv}
\end{align} 
based on our experimental and numerical observations of emergent local randomness during chaotic quantum dynamics (Fig.~\ref{Fig2})~\cite{Supplement}. 

The validity of the relation $F_c \approx F$ can be analytically understood based on the assumption that the projected ensemble of $\ket{\psi_A(z_B)}$ approximately forms a quantum state 2-design. To see this explicitly, we consider
\begin{align} \nonumber
& \sum_{z_B} q(z_B) p_0(z_A|z_B) p (z_A|z_B) \\
\nonumber
&= \sum_{z_B} q(z_B) \bra{\psi_A(z_B)} \hat{P}_{z_A} \ket{\psi_A(z_B)}\bra{\psi_A(z_B)} \hat{V}^\dagger(\tau)\hat{P}_{z_A}\hat{V}(\tau)\ket{\psi_A(z_B)}\\
\nonumber
&= \tr{ \left(\hat{P}_{z_A} \otimes \hat{V}^\dagger(\tau)\hat{P}_{z_A}\hat{V}(\tau) \right) 
\cdot 
\sum_{z_B} q(z_B) (\ket{\psi_A(z_B)} \bra{\psi_A(z_B)})^{\otimes 2}
 }\\
 \nonumber
 &\approx 
\frac{
 \tr{ \left(\hat{P}_{z_A} \otimes \hat{V}^\dagger(\tau)\hat{P}_{z_A}\hat{V}(\tau) \right) 
 \cdot
\left(\hat{\mathds{1}} + \hat{\mathcal{S}}_A \right) }}
{D_A(D_A +1 )}\\
&= \frac{ 1 + \left| \bra{z_A} \hat{V}(\tau) \ket{z_A}\right|^2}
{D_A(D_A + 1)} \label{eq:Fc_equal_F_part}
\end{align}
where $\hat{P}_{z_A} = \ket{z_A}\bra{z_A}$ is the $z$-basis projector onto a specific bitstring $z_A$ in $A$,  $q(z_B)$ is the probability weighting factor, $\hat{\mathds{1}}$ is the identity operator, and $\hat{\mathcal{S}}_A$ is the swap operator acting on subsystem $A$ for the duplicated Hilbert space $\mathcal{H}_A^{\otimes 2}$. In order to obtain the fourth line, we used 
\begin{align}
	\sum_{z_B} q(z_B) (\ket{\psi_A(z_B)} \bra{\psi_A(z_B)})^{\otimes 2} \approx \frac{ \hat{\mathds{1}} + \hat{\mathcal{S}}_A}{D_A (D_A+1)}, \label{eq:2design}
\end{align}
where the right-hand side is due to the projected ensemble forming an approximate quantum state 2-design~\cite{Supplement}. We note that the weighting factors $q(z_B)$ are different than those used for the majority of the manuscript; however, we numerically find that approximate 2-designs form regardless of which weighting factor is used~\cite{Supplement}. 

Inserting Eq.~(\ref{eq:Fc_equal_F_part}) into Eq.~\eqref{eq:Fc_deriv_full}, we obtain
\begin{align}
\nonumber
F_c &\approx \frac{1}{D_A} \sum_{z_A} \left| \bra{z_A} \hat{V}(\tau) \ket{z_A} \right|^2  \\
&= \frac{1}{D} \sum_z \left| \bra{z} \hat{V}(\tau) \ket{z} \right|^2,
\label{method_eqn:Fc_ultimatum}
\end{align}
where the equality on the second line holds because one can always multiply the identity $\frac{1}{D_B} \sum_{z_B} \langle z_B | z_B \rangle^2 = 1$ with the Hilbert space dimension of the complement $D_B= D/D_A$ with $D$ being the Hilbert space dimension of the entire system.

The relation in Eq.~\eqref{method_eqn:Fc_ultimatum} explains how $F_c$ estimates the many-body fidelity with a good accuracy. The right-hand side of Eq.~\eqref{method_eqn:Fc_ultimatum} describes the return probability of $\hat{V}(\tau)$ (also known as Loschmidt echo) averaged over all possible initial states in the fixed measurement basis. Under chaotic time evolution, the propagated error operator $\hat{V}(\tau)$ becomes scrambled, and it is exponentially unlikely in the size of $A$ that a computational state remains unchanged.

Therefore, non-vanishing contributions to $F_c$ arise only when the error operator is partly proportional to the identity, e.g. $\hat{V}(\tau) = c_0 \hat{\mathds{1}} + \sum_s c_s(\tau) \hat{\sigma}_s$ with $c_0 \neq 0$, where $s$ enumerates over all possible Pauli strings supported in $A$. In such a case, $F_c \approx |c_0|^2$ approximates the probability that $\hat{V}$ did not affect the many-body wavefunction, hence $F_c \approx F$. This statement becomes exact if the local qubit on which the error occurs is maximally entangled with the rest of the system at the time of the error. Our analysis can be straightforwardly generalized to more than one error, either located nearby or distant, as long as their joint support $A$ leads to a random ensemble approximately close to the state 2-design.

Finally, we comment on the conditions in which $F_c$ may significantly deviate from $F$. If $\hat{V}$ is diagonal in the measurement basis, e.g. dephasing error along the $z$ axis, and if the error occurs shortly before the bitstring measurements, the return probability in Eq.~\eqref{method_eqn:Fc_ultimatum} will be close to unity despite that the many-body fidelity may be decreased significantly. Our method can fail in this special case. However, if $F_c$ is evaluated after some delay time from the error, then $\hat{V}(\tau)$ becomes scrambled in the operator basis, and $F$ can be approximately estimated (Ext. Data Fig.~\ref{EFig_detection}). In other words, even in the case of the diagonal errors, our formula becomes valid after a finite delay time.

\subsection*{Statistical error scaling from a finite number of bitstring samples}
We quantify the typical statistical error from approximating our fidelity estimator via Eq.~\ref{eq:approxFc} in two steps. First, we use our \textit{ab initio} error model to simulate the quantum evolution of the Rydberg Hamiltonian for system sizes from $N=10$ to 22, from which we can calculate the exact value of $F_c$ compared to error-free numerics. We then sample a finite number of $M$ samples from the probability distributions produced from the error model simulation, apply Eq.~\ref{eq:approxFc}, and plot the standard deviation of $F_c$ as a function of $M$ (Ext. Data Fig.~\ref{EFig_sampling}a). We see a characteristic scaling of $\sigma(F_c) = A/\sqrt{M}$, where $A$ is the \textit{sample complexity}, expected to scale exponentially with $N$, and $\sigma$ denotes the standard deviation. We perform a similar process directly on our experimental data by repeatedly subsampling the experimentally measured bitstrings in order to estimate the scaling of the standard deviation. We plot both error model and experimental results in Ext. Data Fig.~\ref{EFig_sampling}b as a function of $N$, where the time is set as the $N$-dependent entanglement saturation time. By fitting the experimental (error model) results, we find $A\approx a^N$, with $a=1.037(2)$ ($a=1.039(2)$).

\subsection*{Predicting fidelity at the entanglement time}
\noindent\textit{Calculating the entanglement time}---
As can be seen in Fig.~\ref{Fig4}c, entanglement growth in our Rydberg quantum simulator is generally characterized by two distinct regions: a size-independent linear increase, followed by saturation at an $N$-dependent value. In order to systematically capture this behavior and predict the entanglement saturation time for arbitrary $N$, we apply the following protocol. We first calculate the entanglement growth for system sizes ranging from $N=10$ to 22. We then fit all profiles with a functional form of
\begin{align*}
S_\text{ent}(N,t)=\displaystyle\begin{cases} 
      m_1t & t\leq t_c(N) \\
      m_1t_c(N)+m_2(N)(t-t_c(N)) & t>t_c(N)
   \end{cases}
\end{align*}
with free parameters $m_1, m_2(N),$ and $t_c(N)$, but with the explicit constraint that $m_1$ must be the same for all system sizes. From this we find $t_c(N)$, and we further define $t_\text{ent}(N)=Ct_c(N)$, where in the Rydberg case we set $C=1.35$ in order to make sure the time we study is firmly in the saturated entanglement regime (as can be verified visually in Fig.~\ref{Fig4}c). The secondary slope $m_2$ is used because even past $t_\text{ent}$ there is still some slight growth to the entanglement entropy, which becomes more noticeable for larger $N$. This behavior is attributed to slow coupling to the non-blockaded Hilbert space as the blockade constraint is only approximate. The entanglement saturation time is then fit as a linear function of system size, yielding $t_\text{ent}(N) =\alpha_0+\alpha_1 N$; for our particular Hamiltonian parameters we find $\alpha_0=-0.0580(2)$, $\alpha_1=0.05404(1)$, both in units of $\mu$s.

For the case of finding the entanglement saturation depth, $d_\text{ent}$, for the case of random unitary circuits (RUCs) considered in Ext. Data Fig.~\ref{EFig_comparison}, we apply essentially the same procedure. We study two different digital circuit implementations. In the first, with a gate-set based on Ref.~\cite{Arute2019}, the RUC circuit is composed of alternating one- and two-qubit gates; the one-qubit gates are randomly chosen $\pi/2$ rotations along the $\hat{x}$, $\hat{y}$ and $\hat{x}+\hat{y}$ directions, while the two-qubit gates are `fSym'~\cite{Arute2019}. In the second, the RUC is composed entirely of two-qubit SU(4) gates (without global swap operations~\cite{Cross2019}). For the first implementation, we set $C=1$, while in the second we set $C=1.7$, to better guarantee the chosen depth is in the saturated entanglement regime. Open boundary conditions are used in accordance with the experimental Rydberg system, and thus there are two possible gate topologies (i.e. in the first depth applying the `fSym' gate to qubits 1-2, 3-4, etc. or 2-3, 4-5, etc.) - we explicitly average over an equal number of randomized realizations of each topology when calculating the entanglement entropy growth. As in the Rydberg case, we fit the entanglement saturation depth as a linear function of the number of qubits in the RUC, yielding $d_\text{ent}=\beta_0+\beta_1 N_\textrm{RUC}$. For the gate-set based on Ref.~\cite{Arute2019} we find $\beta_0=-0.395(17)$, and $\beta_1=0.557(1)$, and for the gate-set based on SU(4) gates~\cite{Cross2019} we find $\beta_0=-3.18(77)$, and $\beta_1=2.261(51)$.\newline

\noindent\textit{Estimating fidelity decay}---
In Fig.~\ref{Fig4}b, we see the decay profile of the model fidelity, $F_\text{model}$, for our Rydberg simulator is approximately exponential, which we confirm via error model simulations with system sizes ranging from $N=10$ to 22 in Ext. Data Fig.~\ref{EFig_fidelitydecay}a. For each system size, we fit the fidelity decay profile as, 
\begin{align}
\label{eq:ryd_decay}
F(N,t)\propto \mathrm{exp}(-\gamma(N)t), 
\end{align}
where $\gamma(N)$ is the fidelity decay rate. We find that for the system size range considered here, $\gamma(N)$ scales approximately linearly with $N$, from which we fit $\gamma(N)=\gamma_0+\gamma_1 N$; for our particular Hamiltonian parameters and noise sources, we find $\gamma_0=0.12(4)$, and $\gamma_1=0.017(3)$, both in MHz (Ext. Data Fig.~\ref{EFig_fidelitydecay}b). 

In Fig.~\ref{Fig4}d, we use the fitted $\gamma(N)$ explicitly to predict the fidelity scaling of our Rydberg simulator at the $N$-dependent entanglement saturation time, $t_\mathrm{ent}$, as a function of system size. Concretely, we plot (red dashed line):
\begin{align}
\label{eq:ryd_decay_ent}
F_\text{model}(N, t_\text{ent}(N)) = F_0^N \mathrm{exp}(-\gamma(N)t_\text{ent}(N)),
\end{align}
where $F_0{=}0.997(1)$ is the single-atom preparation fidelity determined experimentally. The shaded red region in Fig.~\ref{Fig4}d depicts the error from fit uncertainty of $\gamma(N)$.

For the RUC case, fidelity decay for a given system size, $N_\textrm{RUC}$, and depth, $d$, is modeled as a simple product over constituent two-qubit cycle fidelity, $F_{\mathrm{cycle}}$, yielding
\begin{align}
\label{eq:ruc_decay}
F_\mathrm{RUC}(N,d)=F_{\mathrm{cycle}}^{(N_\textrm{RUC}-1)d/2},
\end{align}
where the exponent of the right-hand side reflects the fact that we apply, on average, $(N_\textrm{RUC}-1)/2$ two-qubit gates in parallel per depth.\newline

\noindent\textit{Comparing digital and analog devices}---
We wish to directly compare the evolution fidelity of our analog Rydberg quantum simulator against that of a digital device implementing an RUC with equivalent entanglement entropy at the entanglement saturation time. However, due to the Rydberg blockade mechanism, as well as symmetries of our Hamiltonian~\cite{Supplement}, an equal number of atoms in the Rydberg simulator, $N$, and qubits in the RUC, $N_\textrm{RUC}$, will not saturate to the same half-chain entanglement entropy. 

To overcome this, in Ext. Data Fig.~\ref{EFig_comparison}b we plot the entanglement entropy, $S$ ($S_\mathrm{RUC}$), achieved at $t_\text{ent}$ ($d_\text{ent}$) for the Rydberg simulator (RUC) as a function of $N$ ($N_\textrm{RUC}$). For the Rydberg simulator, we fit $S(N)=\sigma_0+\sigma_1 N$ with $\sigma_0=0.16(4), \sigma_1=0.26(3)$. For the RUC, we use the prediction of $S_\textrm{RUC}(N_\textrm{RUC}) = \eta_0+\eta_1 N_\textrm{RUC}$ with $\eta_0=-\mathrm{log}_2(e)/2\approx-0.72$ and $\eta_1=1/2$ being exact values with no error bars, as defined in Ref.~\cite{Page1993} (where $e$ is Euler's number, and where we have used the $\mathrm{log}_2$ entanglement entropy convention). To find the equivalent $N_\textrm{RUC}$ for a given $N$, we then simply equate $S_\textrm{RUC}(N_\textrm{RUC})=S(N)$, yielding $N_\textrm{RUC}=(\sigma_1 N + (\sigma_0-\eta_0))/\eta_1=0.52 N + 1.76$. 

With this system size equivalence established we can directly compare the SPAM-corrected Rydberg and RUC systems at their respective entanglement time and depth, in order to find the equivalent RUC two-qubit cycle fidelity which would match the Rydberg quantum simulator's evolution fidelity. By evolution fidelity, we refer to the fidelity at the entanglement time, up to preparation errors, which based on our validated error model is approximately given by $\mathrm{exp}(-\gamma(N)t_\text{ent}(N))$ from Eq.~\ref{eq:ryd_decay_ent}. We equate $F_\mathrm{RUC}(N_\textrm{RUC},d_\text{ent})=\mathrm{exp}(-\gamma(N)t_\text{ent}(N))$, and then solve for $F_{\mathrm{cycle}}$. As shown in Ext. Data Fig.~\ref{EFig_comparison}c, for the gate-set used in Ref.~\cite{Arute2019}, we find $F_{\mathrm{cycle}}=0.987(2)$, while for the SU(4) circuit we find $F_{\mathrm{cycle}}=0.9965(5)$, nearly independent of system size. Error bars originate from the uncertainty on the parameters of $\gamma$, $t_\mathrm{ent}$, $d_\text{ent}$, $F_0$, and $S$.

\subsection*{Target state benchmarking}
Our fidelity estimation protocol can be used both to estimate the fidelity of performing some quantum evolution (Fig.~\ref{Fig4}), and to estimate the fidelity of \textit{preparing} a target quantum state of interest (Fig.~\ref{Fig5}). In this modality, we assume the target state is prepared with some non-unity fidelity due to experimental imperfections, after which we apply an infinite effective temperature quench Hamiltonian, and observe the resulting dynamics. 

In Fig.~\ref{Fig5}d, the ideal state is the ground state at $\Delta/\Omega=3$, $V_\text{nnn}/\Omega=0.26$, close to the phase transition between the disordered and $\mathbb{Z}_2$-ordered states of the Rydberg Hamiltonian~\cite{Slagle2021}. The imperfect state is taken to be an incoherent mixture composed of 50\% each of the ground and first excited states. This state is then quenched with a Hamiltonian with parameters $\Omega/2\pi=5.3$ MHz, $\Delta/2\pi=2.8$ MHz, $C_6/2\pi=254$ GHz $\mu$m$^6$, $a=3.75~\mu$m, with $2\pi\times\pm1$MHz random on-site disorder drawn from a uniform distribution.

In Ext. Data. Fig.~\ref{EFig_tsb}, we numerically demonstrate fidelity estimation of various target states such as a cluster state, a Haar-random state of a 2D Rydberg quantum simulator, and a symmetry-protected topological (SPT) ground state.

Specifically, in Ext. Data. Fig.~\ref{EFig_tsb}a, we estimate the state preparation fidelity of a one-dimensional cluster state defined as
\begin{align}
	\ket{\psi}_\text{cluster} = \prod_{i=1}^{N-1} \text{(CZ)}_{i,i+1} \ket{+}^{\otimes N}
\end{align}
where $\text{(CZ)}_{i,i+1}$ is the two-qubit, controlled-$Z$ gate acting on two adjacent qubits at site $i$ and $i+1$, and $\ket{+}$ is the equal superposition of the $\ket{0}$ and $\ket{1}$ states. The imperfect quantum state is prepared by applying a global phase rotation to the ideal state such that the state overlap becomes $F = 0.5$. We then employ an infinite effective temperature quench Hamiltonian given as $\hat{H}/\hbar = h_x \sum_i (\hat{S}_i^x - 1.79 \hat{S}_i^y + 4.64 \hat{S}_i^x \hat{S}_{i+1}^x)$ to learn the state overlap via our $F_c$ formula. 

In Ext. Data. Fig.~\ref{EFig_tsb}b,  we estimate the state preparation fidelity of a pure Haar-random state generated from Rydberg atoms in a $4\times4$ two-dimensional square array. The imperfect state is prepared by applying a local phase rotation to a central qubit, yielding $F = 0.5$. For subsequent quench dynamics, identical Hamiltonian parameters are used as in the 1D Rydberg benchmarking case (Fig.~\ref{Fig4}).

Lastly, in Ext. Data. Fig.~\ref{EFig_tsb}c, we estimate the state preparation fidelity of a symmetry-protected topological (SPT) ground state prepared in a Rydberg ladder array realizing the Su-Schrieffer-Heeger topological model, following the approach of Ref.~\cite{de2019}. The imperfect state is prepared with a local phase error yielding a state overlap of $F = 0.5$. It is subsequently benchmarked via infinite effective temperature evolution with the quench Hamiltonian chosen to be the combination of the identical interaction Hamiltonian, a random on-site disorder of strength 1 MHz, and a detuned global drive with a Rabi frequency of 2 MHz and a detuning of 0.5 MHz. 

\section*{Acknowledgements}
We acknowledge help from Pascal Scholl during the revision of this manuscript as well as discussions with Abhinav Deshpande and Alexey Gorshkov. We acknowledge funding from the Institute for Quantum Information and Matter, an NSF Physics Frontiers Center (NSF Grant PHY-1733907), the NSF CAREER award (1753386), the AFOSR YIP (FA9550-19-1-0044), the DARPA ONISQ program (W911NF2010021), the Army Research Office MURI program (W911NF2010136), the NSF QLCI program (2016245), the DOE (DE-SC0021951) and Fred Blum. JC acknowledges support from the IQIM postdoctoral fellowship. ALS acknowledges support from the Eddleman Quantum graduate fellowship. RF acknowledges support from the Troesh postdoctoral fellowship. JPC acknowledges support from the PMA Prize postdoctoral fellowship. HP acknowledges support by the Gordon and Betty Moore Foundation. HH is supported by the J. Yang \& Family Foundation. AK acknowledges funding from the Harvard Quantum Initiative (HQI) graduate fellowship. JSC is supported by a Junior Fellowship from the Harvard Society of Fellows and the U.S. Department of Energy under grant Contract Number DE-SC0012567. SC acknowledges support from the Miller Institute for Basic Research in Science. 

\end{document}


\title{Supplementary information: Preparing random states and benchmarking with many-body quantum chaos}

\author{Joonhee Choi}\thanks{These authors contributed equally to this work}
\author{Adam L. Shaw}\thanks{These authors contributed equally to this work}
\author{Ivaylo S. Madjarov}
\author{Xin Xie}
\author{Ran Finkelstein}
\affiliation{\Caltech}
\author{\\Jacob P. Covey}
\affiliation{\Caltech}
\affiliation{\UIUC}
\author{Jordan S. Cotler}
\affiliation{\Harvard}
\author{Daniel K. Mark}
\affiliation{\MIT}
\author{Hsin-Yuan Huang}
\affiliation{\Caltech}
\author{Anant Kale}
\affiliation{\Harvard}
\author{\\Hannes Pichler}
\affiliation{\Innsbruck}
\affiliation{\InnsbruckCenter}
\author{Fernando G.S.L. Brand{\~{a}}o}
\affiliation{\Caltech}
\author{Soonwon Choi}\email{soonwon@mit.edu}
\affiliation{\MIT}
\affiliation{\Berkeley}
\author{Manuel Endres}\email{mendres@caltech.edu}
\affiliation{\Caltech}

\maketitle
\tableofcontents
\pagebreak

\section{Projected Ensemble and Comparison to Quantum State $k$-designs}

To characterize the randomness of the projected ensemble, we compare it to the Haar-random ensemble which represents an ensemble of uniformly random pure states defined in the Hilbert space of subsystem $A$. Specifically, the similarity between the two ensembles can be characterized via the $\ell_1$-norm that measures a `distance' in operator space (referred to as the `trace distance' in the main text),
\begin{align}
	\ell_1^{(k)} = \Vert \rho^{(k)}_A - \rho^{(k)}_\text{Haar} \Vert_1. \label{eq:L1norm}
\end{align}
Here $ \Vert \cdot \Vert_1 $ represents the absolute sum of singular values, $\rho_A^{(k)}$ is the $k$th moment of the projected ensemble consisting of local states $\ket{\psi_A(z_B)}$ in subsystem $A$

\begin{align}
\rho_A^{(k)} = \sum_{z_B} p(z_B) (\ket{\psi_A(z_B)} \bra{\psi_A(z_B)})^{\otimes k},
\end{align}
and $\rho^{(k)}_\text{Haar}$ is the $k$th moment of the Haar-random ensemble 
\begin{align}
	\rho_{\text{Haar}}^{(k)} = \int_{\text{Haar($D_A$)}} d\psi \, \left(|\psi\rangle \langle \psi|\right)^{\otimes k}. \label{eq:rho_Haar}	 
\end{align}
From the Schur-Weyl duality, Eq.~(\ref{eq:rho_Haar}) can be simplified to
\begin{equation}
\label{E:keyidentity1}
\rho^{(k)}_\text{Haar} = \frac{\sum_{\pi \in \mathcal{S}_k} \text{Perm}_{\mathcal{H}_A^{\otimes k}}(\pi)}{D_A(D_A+1)\cdots (D_A+k-1)},
\end{equation}
where $\text{Perm}_{\mathcal{H}_A^{\otimes k}}(\pi)$ permutes the $k$ copies of the Hilbert space $\mathcal{H}_A$ of subsystem $A$ according to a member, $\pi$, in the permutation group of $k$ elements, $ \mathcal{S}_k$, and $D_A$ is the dimension of $\mathcal{H}_A$~\cite{Harrow2013}.

\begin{figure}[t!]
	\centering
	\includegraphics[width=89mm]{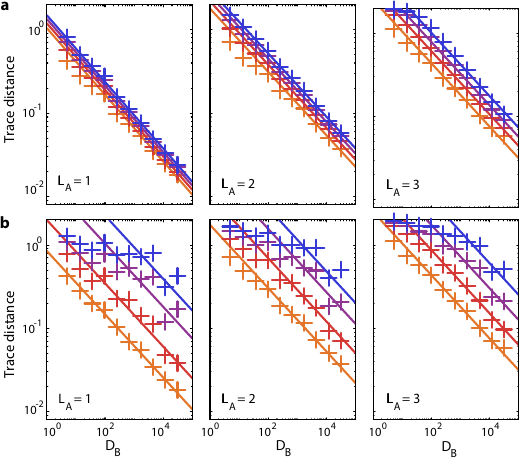}
	\caption{\textbf{Trace distance between the projected ensemble and quantum state $k$-designs: } \textbf{a, b,} System size scaling of late-time trace distances from three different subsystem sizes of $L_A = 1,2,$ and 3. For fixed subsystem $A$ size, we increase the Hilbert space dimension of outer subsystem $B$, $D_B$, and compute trace distances to the quantum state $k$-designs for four different orders, $k=1$ (orange), 2 (red), 3 (purple), and 4 (blue). Sampling weighting factors of $p(z_B)$ and $p^k(z_B) / \sum_{z_B} p^k(z_B)$ are used in \textbf{a} and \textbf{b}, respectively, in the construction of a $k$th-order projected ensemble.} \vspace{-0.5cm}
	\label{EFig_distance}
\end{figure}

As shown in Fig.~3 and Supp. Fig.~\ref{EFig_distance}, we find that for various subsystem lengths, $L_A$, the $\ell_1$-norm distance decreases for all orders $k=1,2,3,$ and $4$ with increasing Hilbert space dimension, $D_B$, of subsystem $B$. Non-zero $\ell_1$-norm distances, as well as statistical fluctuations observed in the moments at {\it finite} $D_B$, are associated with a finite sample size effect in the constructed projected ensemble. For a finite-size Haar-random ensemble, the trace distance to the uniform ensemble scales as ${\sim}1/\sqrt{D_B}$, consistent with the scaling result of the projected ensemble~\cite{Cotler2021}. These results are obtained with $\Omega/2\pi = 4.7$ MHz, $\Delta/2\pi = 0.5$ MHz, $C_6 = 2\pi \times 254$ GHz $\mu \textrm{m}^6$, and $a=3.75$ $\mu$m; the discrepancy between this $\Delta$ and that imposed in experiment does not result in a significantly different distance to the $k$-designs for the experimental system size of ten qubits.

\begin{figure}[t!]
	\centering
	\includegraphics[width=89mm]{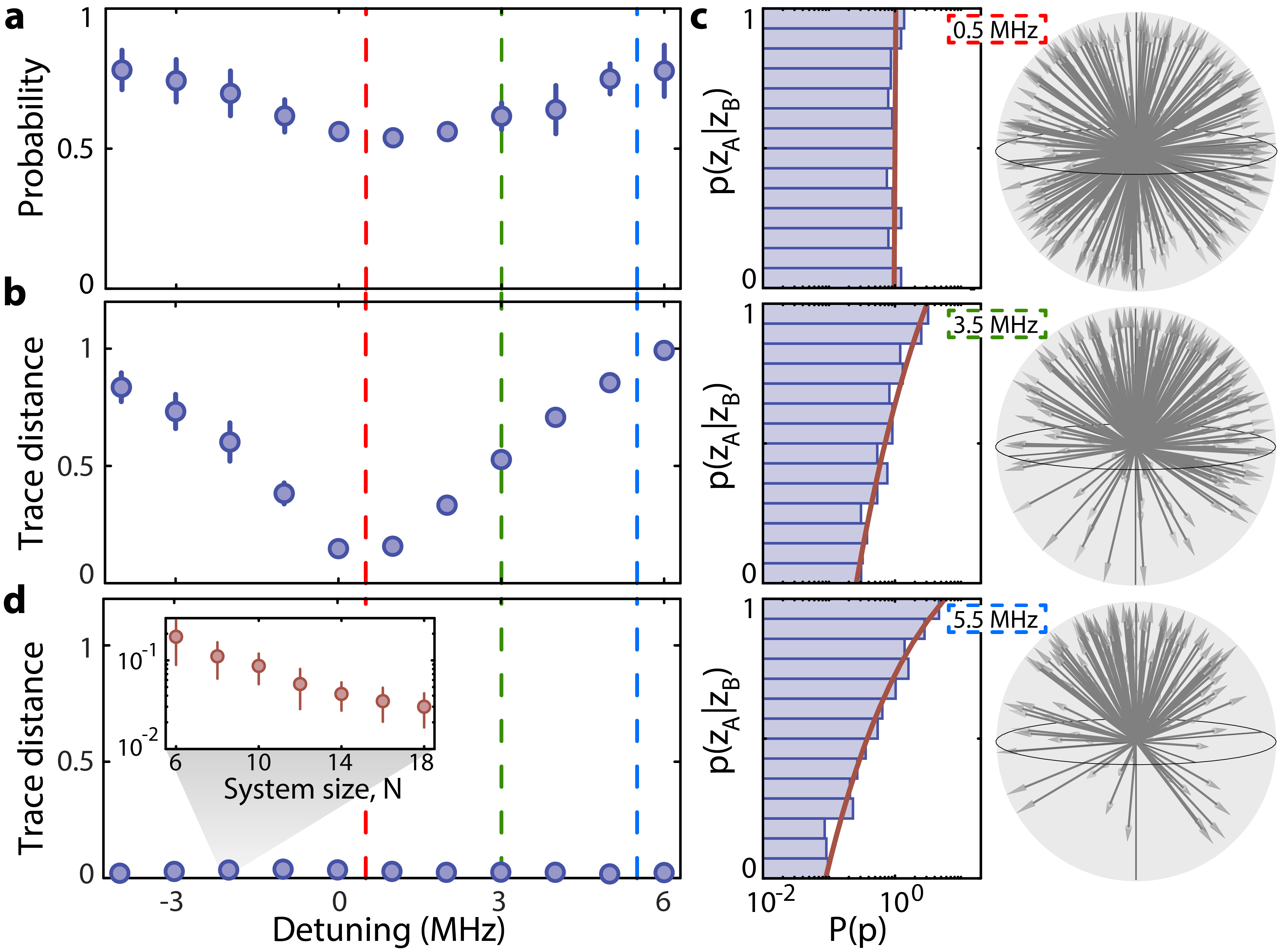}
	\caption{\textbf{Signatures of finite effective temperature projected ensembles. a,} Late-time, marginal probability values of $p(z_A=0)$ as a function of global detuning for the Rydberg Hamiltonian with $N=16$ atoms; error bars denote the standard deviation over temporal fluctuations. We find that at a detuning value of 0.5 MHz, the late-time probability value equilibrates around 0.5, signaling the infinite effective temperature condition where a uniformly random ensemble emerges. \textbf{b,} Trace distances between the projected ensemble and the Haar-random ensemble as a function of the global detuning; as in \textbf{a}, the distance is minimized at the infinite effective temperature condition of 0.5 MHz. \textbf{c,} Projected ensembles for a single-qubit subsystem and their conditional probability distributions at finite effective temperatures. Detuning values are chosen as 0.5, 3, and 5.5 MHz (indicated by dashed lines in \textbf{a} and \textbf{b}), respectively from top to bottom. Here, the subsystem $A$ is taken from the middle atom at site 8. We find the numerical results are in good agreement with the analytic prediction from the so-called \textit{Scrooge ensemble} (red lines, see text). \textbf{d,} Trace distances between the projected ensemble and the Scrooge ensemble as a function of the global detuning. The inset shows that the trace distance decreases monotonically as increasing total system size, $N$, which suggests that the single-qubit, finite-temperature projected ensembles converge to the Scrooge ensemble in the thermodynamic limit.} \vspace{-0.5cm}
	\label{EFig_thermal}
\end{figure} 

\section{Signatures of finite effective temperature projected ensembles}
\label{sec:thermal}

While all measurements in the main text were conducted with the system thermalizing close to an infinite effective temperature condition, it is interesting to ask how the emergent randomness of the projected ensemble is changed in the finite effective temperature case. To this end, we numerically simulate quench evolution under the Rydberg Hamiltonian with different choices of global detuning, and observe the effect on observables for an $N=16$ atom system. In Supp. Fig.~\ref{EFig_thermal}a, we first show the late-time marginal probability of measuring a single-qubit subsystem $A$ in the state $\ket{0}$, which is minimized at a detuning of 0.5 MHz, with a probability close to 0.5; this condition corresponds to the infinite effective temperature case. Similarly, we find the distance between the projected ensemble and the Haar-random ensemble (at the level of a 2nd order moment) also is minimized at this point (Supp. Fig.~\ref{EFig_thermal}b). For three choices of detuning, indicated by vertical dashed lines in Supp. Fig.~\ref{EFig_thermal}a and b, we plot the probability histograms for the single qubit subsystem, as in Fig. 2 of the main text, and show the corresponding distribution of projected ensemble state vectors on the Bloch sphere (Supp. Fig.~\ref{EFig_thermal}c).

We find that the numerical data in Supp. Fig.~\ref{EFig_thermal}c agree well with an analytical form, $P_S(p)=(a p+b (1-p))^{-3}$, where $a$ and $b$ are real constants fixed such that the distribution is normalized, $\int_0^1 dp P_S(p) =1$, and the mean of the distribution corresponds to the marginal probability, $\int_0^1 dp P_S(p) p = p(z_A = 1)$ (red curves). This distribution, $P_S(p)$, is predicted from a so-called Scrooge ensemble of states, first proposed in Ref.~\cite{Jozsa1994} in the context of \textit{quantum communication} (see Ref.~\cite{Goldstein2006} for a derivation of $P_S(p)$). It was later hypothesized that such an ensemble could appear in quantum chaotic dynamics~\cite{Goldstein2016}. The Scrooge ensemble consists of normalized states, $\ket{\psi}_\rho=\hat \rho^{1/2}\ket{\psi} /\sqrt{\bra{\psi}\hat\rho\ket{\psi}}$, where $\ket{\psi}$ is drawn from a uniform, Haar-random ensemble with a weighing factor $D_S \bra{\psi}\hat{\rho}\ket{\psi}$, where $D_S$ is the dimension of the Hilbert space in which the Scrooge ensemble is defined. As its first-order moment, the Scrooge ensemble reproduces a desired target density operator, $\hat \rho$:  $E_{\psi \in Haar}[D_S \bra{\psi}\hat{\rho}\ket{\psi} \ket{\psi}_\rho\bra{\psi}_\rho]=\hat \rho$; in our case, $\hat \rho$ refers to the reduced density operator of $A$. For the effective infinite effective temperature case ($\hat \rho=\hat I/D_A$), the Scrooge ensemble reduces to a homogeneous, Haar-random ensemble. Motivated by the the good agreement of probability histograms (Supp. Fig.~\ref{EFig_thermal}c), we also plot the trace distance between the second moment of the Scrooge ensemble, $E_{\psi \in \text{Haar}}[D_S \bra{\psi}\hat{\rho}\ket{\psi} (\ket{\psi}_\rho\bra{\psi}_\rho)^{\otimes 2}]$, and the second moment of the projected ensemble as a function of detuning (Supp. Fig.~\ref{EFig_thermal}d). In contrast to the Haar ensemble,  the distance to the Scrooge ensemble is now almost independent of detuning. Moreover, for a detuning that corresponds to finite effective temperature, we observe that the distance decreases monotonically with total system size (inset of Supp. Fig.~\ref{EFig_thermal}d).\\
\indent It is interesting to note that the projected ensemble can indeed be viewed through a quantum communication lens: projective measurements in $B$ create the projected ensemble, which can be thought of as a message ensemble sent to an observer in $A$. Message ensembles are often characterized by the accessible information, the maximum information gain the observer (by optimal measurement) can achieve about the message. Interestingly, Scrooge ensembles attain a lower bound of the accessible information if used as a message ensemble~\cite{Jozsa1994}. Our results in Supp. Fig.~\ref{EFig_thermal} suggest that Scrooge ensembles naturally appear in quantum many-body dynamics as projected ensembles (from a generic local basis measurement in $B$), hence minimizing the accessible information about the projected ensemble obtainable from measurements in $A$. We leave a more detailed analysis of this phenomenon to future work.

\section{Moments and Two-point correlations of Uniformly Random Ensembles}

\textbf{Moments}.~Suppose we have a Haar-random state $|\psi\rangle$ in a $D_A$-dimensional Hilbert space, which we measure in the standard $z$-basis.  Then the probability of measuring a bitstring $z$ is $r_\psi(z) = |\langle z |\psi\rangle|^2$.  We can compute the expectation value of the $k$th moment of $r_\psi(z)$ over the Haar-random ensemble as
\begin{align}  \label{eq:Haarmoments}
\mathbb{E}_{\psi \in \text{Haar}(D_A)}\!\left[r_\psi(z)^k \right] &= \frac{k!}{D_A(D_A+1) \cdots (D_A +k -1)}\,.
\end{align}
Using these moments, we can construct a probability density function
\begin{equation}
P_H(p) \equiv \mathbb{E}_{\psi \sim \text{Haar}(D_A)}\!\left[ \delta(r_\psi(z) - p)\right]\,
\end{equation}
where $\delta$ is the Dirac delta function, yielding $r_\psi(z) = p$ over the Haar-random ensemble. Since $\int_0^1 dp \, P_H(p) \,p^k = \mathbb{E}_{\psi \sim \text{Haar}(D_A)}\!\left[r_\psi(z)^k \right] = 1/\binom{D_A+k-1}{k}$ for $k=0,1,2,...$, we find
\begin{equation}
P_H(p) = (D_A-1) (1 - p)^{D_A-2}\,.
\end{equation}

\begin{figure}[t!]
	\centering
	\includegraphics[width=89mm]{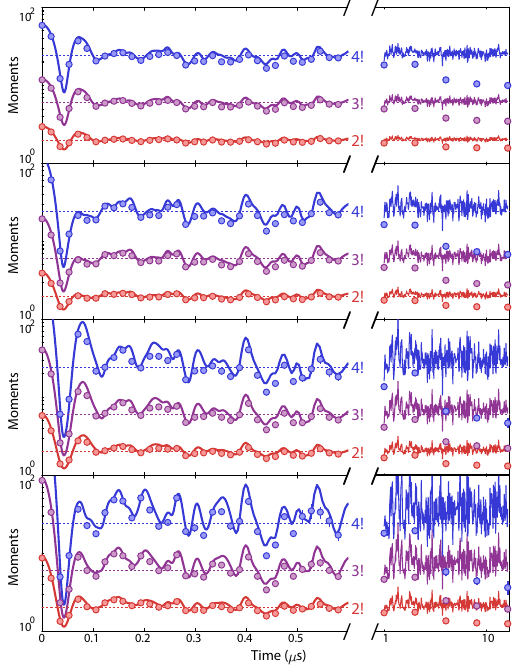}
	\caption{\textbf{Moments of the projected ensemble.} 2nd (red), 3rd (purple), and 4th (blue) moments of the projected ensemble for different choices of the size of $A$, ranging from $L_A=1$ (top) to $L_A=4$ (bottom). Data is taken from the same dataset as in Fig. 3 of the main text.} \vspace{-0.5cm}
	\label{EFig_moments}
\end{figure} 

\begin{figure*}[t!]
	\centering
	\includegraphics[width=183mm]{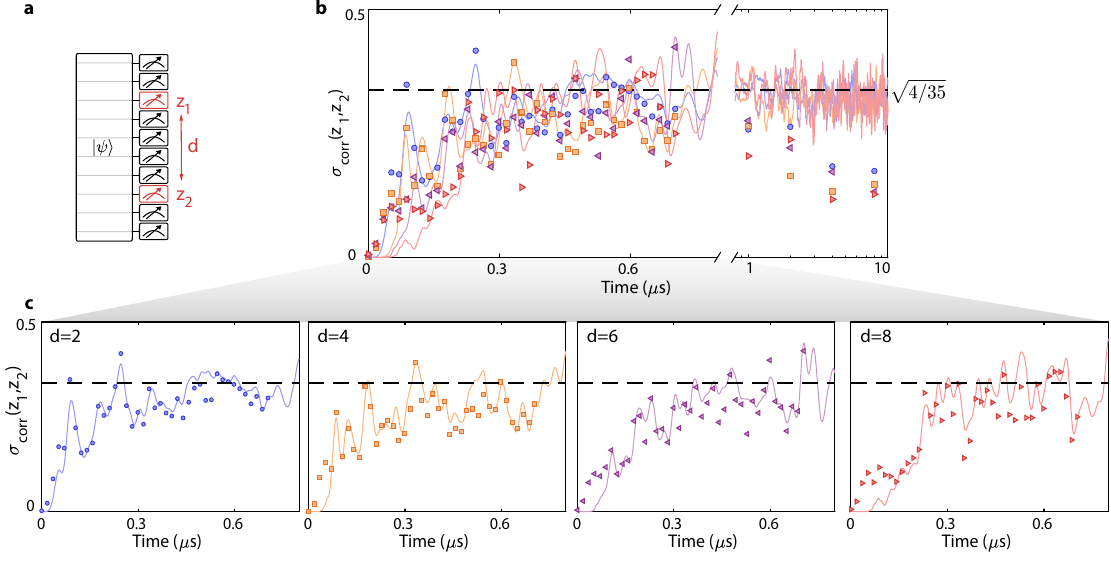}
	\caption{\textbf{Fluctuations of conditional two-point correlations in projected state ensembles: }  \textbf{a,} Two-point correlations based on conditional projected states of two qubits separated at distance $d$. \textbf{b,} Fluctuations of two-point correlation functions of conditional probabilites, $\sigma_\text{corr}(z_1,z_2)$, characterized with a ten-qubit Rydberg-atom array simulator (Eq.~\eqref{eq:corrfunction}). Experimental (markers) and numerical (solid lines) results increase and saturate to near the theoretical prediction of $\sigma_\text{corr} = \sqrt{4/35}$ from the uniformly random ensemble. At late times $t > 1~\mu$s, the universal feature gradually vanishes due to accumulated errors and decoherence processes in the experiment (time scale is logarithmic after the break). Colors and markers used are: $d=2$, blue circle; $d=4$, orange square; $d=6$, purple triangle; $d=8$; red triangle. \textbf{c,} Individual plots for the four different pair separations shown in \textbf{b}.}\vspace{-0.5cm}
	\label{EFig_correlation}
\end{figure*}

\textbf{Correlation functions.}~In Supp. Fig.~\ref{EFig_correlation}, we measure the square root of a weighted average of squared two-point correlators in the $z$-basis using the projected ensemble of a two-qubit subsystem, given as
\begin{align}
	\sigma_\text{corr}(z_1,z_2) = \sqrt{\sum_{z_B} p(z_B) C(z_1, z_2| z_B)^2}
	\label{eq:corrfunction} 
\end{align}
where 
\begin{align}
C(z_1, z_2 | z_B) &= \langle \psi_A(z_B) |\hat{Z}_1 \hat{Z}_2 | \psi_A(z_B) \rangle -  \\ \nonumber
&\langle \psi_A(z_B) | \hat{Z}_1 | \psi_A(z_B) \rangle \langle \psi_A(z_B) | \hat{Z}_2 | \psi_A(z_B) \rangle. 
\end{align}
Here, $\hat{Z}_i$ is the Pauli-$Z$ operator at site $i$, $p(z_B)$ is the probability of observing a bitstring $z_B$ from outer subsystem $B$, and the expectation values on the right-hand side are computed with the projected two-qubit state $\ket{\psi_A(z_B)}$. To compare the projected ensemble result with a Haar-random ensemble, we use Eq.~\eqref{E:keyidentity1} for $D_A=4$ to compute the analytical value of such a correlator for the Haar-random ensemble: 
\begin{align}
\label{E:expHaarcorrZZ2}
\sqrt{\mathbb{E}_{\psi \in \text{Haar}(4)}\!\left[\left(\langle \psi | \hat{Z}_1 \hat{Z}_2 |\psi\rangle - \langle \psi| \hat{Z}_1 |\psi\rangle \langle \psi| \hat{Z}_2 |\psi\rangle\right)^2\right]} = \sqrt{\frac{4}{35}}.
\end{align}

\section{Benchmarking of random unitary circuit and chaotic Hamiltonian}
In Ext.~Data Fig.~3, we present simulated benchmarking results for both digital and analog quantum systems based on a random unitary circuit (RUC) and a trapped ion Hamiltonian exhibiting chaotic behaviors, respectively; below we describe details for simulations of their quantum dynamics in the presence of errors.

\textbf{Random unitary circuit:} The dynamics of the one-dimensional RUC in Ext.~Data Fig.~3 are simulated using random two-qubit SU(4) unitary gates sampled from the Haar measure. Specifically, the time evolution of an $N$-qubit system starting from the all-zero initial state, $|0\rangle^{\otimes N}$, can be described as
\begin{align}
|\psi(t) \rangle = \hat{\mathcal{U}}_t  \hat{\mathcal{U}}_{t-1} \cdots \hat{\mathcal{U}}_{2}  \hat{\mathcal{U}}_1 |0\rangle^{\otimes N}, 
\end{align}
where $\hat{\mathcal{U}}_\text{odd} = \{ \hat{\mathcal{U}}_1, \hat{\mathcal{U}}_3, \cdots \}$ and $\hat{\mathcal{U}}_\text{even} = \{ \hat{\mathcal{U}}_2, \hat{\mathcal{U}}_4, \cdots \}$ are the odd- and even-time unitaries modeled as
\begin{align}
 \hat{\mathcal{U}}_\text{odd} = \prod_{i=1}^{N/2} \hat{U}_{2i-1,2 i}, \quad \hat{\mathcal{U}}_\text{even} &= \prod_{i=1}^{N/2} \hat{U}_{2i,2i+1}
\end{align} 
with open boundary condition, and $t$ is the circuit depth and $\hat{U}_{\mu,\nu}$ is the randomly-sampled SU(4) gate acting on two qubits at site $\mu$ and $\nu$. Note that at each circuit depth $t$, we randomly sample two-qubit random unitaries to realize a many-body unitary $\hat{\mathcal{U}}$ giving rise to chaotic dynamics.

\textbf{Trapped ion Hamiltonian:} 
We simulate a trapped ion Hamiltonian that leads to thermalization of the all-zero initial state to infinite effective temperature~\cite{Cirac1995}: 
\begin{align}
	\hat{H}_\text{ion}/\hbar = 2\pi J \left[\sum_{i=1}^N 0.4 \hat{S}_i^x + 0.45 \hat{S}_i^y + \sum_{i=1}^{N-1} \sum_{j=i+1}^N \frac{\hat{S}_i^x \hat{S}_j^x}{|i-j|} \right], 
\end{align}
where $\hat{S}_i^{x,y,z}$ are the spin-1/2 operators at site $i$. The Hamiltonian coefficients are adapted from Ref.~\cite{Kim2014}, and are consistent with those used in an accompanying theory paper except for the fact that here we include $1/r$ long-range interactions~\cite{Cotler2021}. In addition, coefficients are globally rescaled by $J$ such that the early-time growth rate of half-chain entanglement entropy is matched to that of RUC dynamics. The many-body quantum state at time $t$ is obtained by solving for $|\psi(t)\rangle = e^{-i \hat{H}_\text{ion}t/\hbar} |0\rangle^{\otimes N}$.

\textbf{Noisy time evolution:} To emulate noisy quantum dynamics for the RUC and the trapped ion system, we employ a stochastic evolution method with a simple noise model that considers local bit-flip and phase-flip errors represented by the Pauli operators $\hat{S}^{x,y,z}$. These local errors are stochastically applied to individual qubits at a single-qubit error rate of $\gamma_\text{err}$ per unit time (equivalent to a single circuit depth); the particular error applied is chosen randomly from the set of $\hat{S}^{x,y,z}$. To compare the RUC and Hamiltonian on an equal footing, we run noisy simulations at the same $\gamma_\text{err}$ between the two cases. Repeating the noisy dynamics simulations more than $\sim$10000 times at a fixed $\gamma_\text{err}$, we obtain good approximations of the density matrices, $\hat\rho_\text{RUC}$ and $\hat\rho_\text{Ham}$, from which we extract the true fidelity $F = \langle \psi | \hat\rho | \psi \rangle $ as well as our fidelity estimator $F_c$ and the linear cross-entropy benchmark~\cite{Arute2019}
\begin{align}
	F_\text{XEB} = (D+1) \sum_z p_0(z) p(z) - 1,
\end{align}
where $D=2^N$ is the Hilbert space dimension of a global $N$-qubit system, $z$ is the global bitstring of length $N$, and $p_0$ and $p$ are the bitstring probabilities in the $z$-basis without and with errors, respectively. 

\section{Sensitivity in Hamiltonian parameter estimation}
The Hamiltonian parameter estimation scheme presented in Fig.~5 of the main text allows for estimating both global and local Hamiltonian parameters; here we show how its sensitivity compares to, for instance, a metric based on the residual-sum-of-squares between the local magnetization from experiment (or any noisy reference) and from error free numerics,
\begin{align}
RSS(\Omega,t) = \sum\limits_{i=1}^N(\langle \hat{S}_z^i(\Omega)\rangle-\langle \hat{S}_{z,\text{exp}}^i\rangle)^2,
\end{align}
where $\hat{S}_z^i$ the local magnetization along the $z$ axis for the atom at site $i$ from the numerical simulation, and $\hat{S}_{z,\text{exp}}^i$ is the same magnetization from a fixed reference, for instance the experimental data. Here, $\Omega$ can be regarded as {\it any} parameter that determines the Hamiltonian of interest (e.g. Rabi frequency). We hypothesize that while the $RSS$ will be peaked at the correct set of Hamiltonian parameters, at late times it will be less sensitive than our proposed protocol based on $F_c$ because while the local magnetization will thermalize to a fixed value, the global state, which $F_c$ is sensitive to, will still be continuously evolving.

We test this hypothesis via error-free numerical simulation: first we simulate the Rydberg Hamiltonian with the Rabi frequency fixed at $\Omega/2\pi=5.3$ MHz - this set of numerics serves as the reference. Next, we repeat the simulation several times, each time with a different Rabi frequency, varying from $\Omega/2\pi=5.15$ MHz to $5.45$ MHz. We then compare these simulations on the level of $F_c$ and $RSS$ to identify how well both estimation methods can identify $5.3$ MHz as the correct Rabi frequency in the original simulation.

\begin{figure}[t!]
	\centering
	\includegraphics[width=89mm]{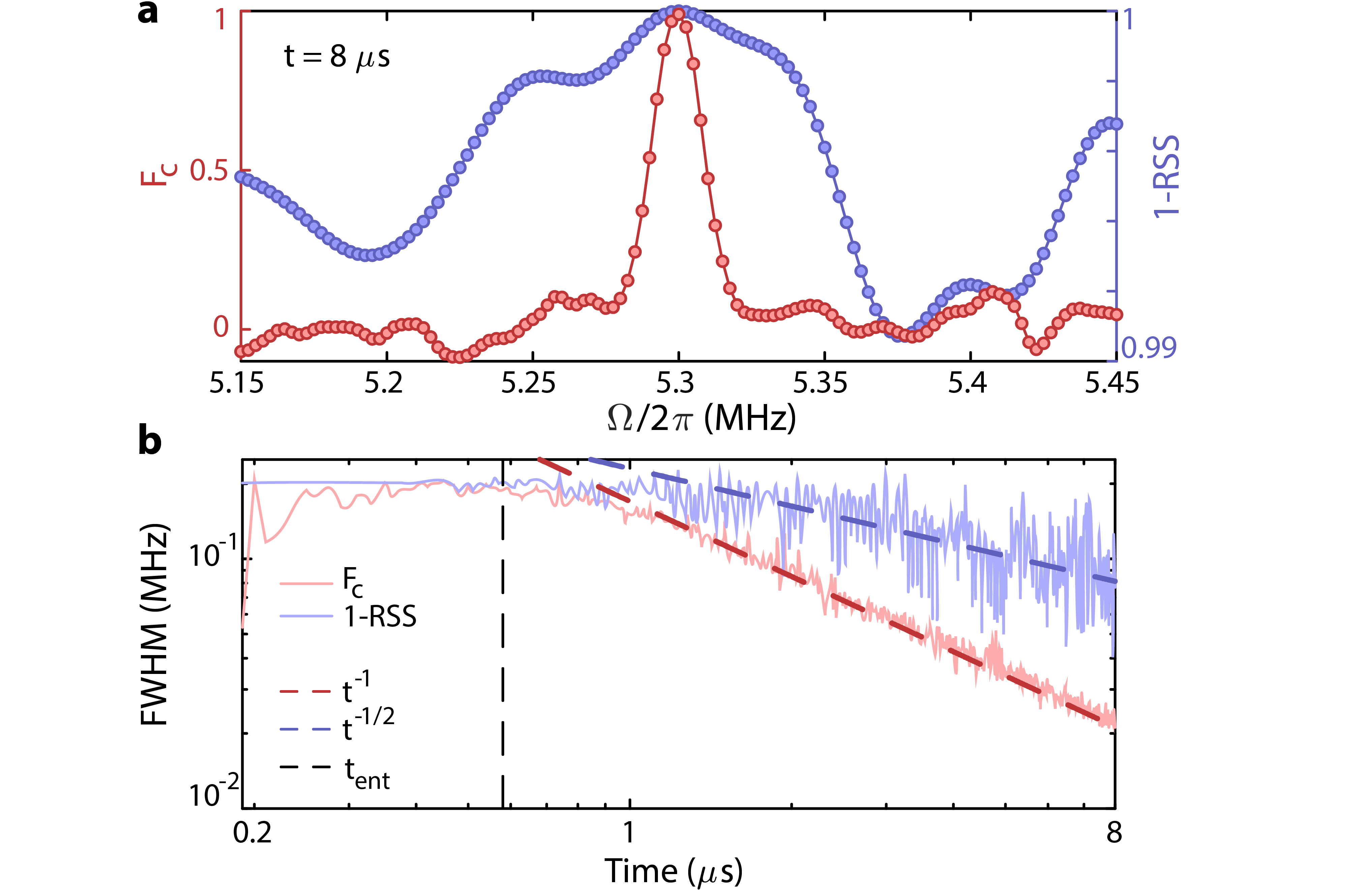}
	\caption{\textbf{Sensitivity of Hamiltonian parameter estimation. a,} We perform an error-free numerical simulation of the Rydberg Hamiltonian with $N=12$ and fixed Rabi frequency, $\Omega=5.3~\text{MHz}$, which is then compared against another error-free simulation with varying $\Omega$. To explore the efficacy of our parameter estimation scheme, we compare the two simulations in two ways: via $F_c$ (red, as in Fig. 5a of the main text), and via $1-RSS$, where $RSS$ is the residual-sum-of-squares of the local magnetizations (blue, see text); both estimators are taken at a single time point of $t=8 ~\mu$s. We note that $F_c$ is both more sharply peaked at the correct $\Omega$, and that the range of values which $F_c$ takes is larger than for $1-RSS$, which has a dynamic range of only 0.01. \textbf{b,} The FWHM of the central peak in $F_c$ in \textbf{a} stays relatively flat up until $t_{\mathrm{ent}}$, the entanglement time of the system (see Fig. 4d of the main text for example), before decreasing as $1/t$ at late times. The FWHM of the RSS estimator follow similar behavior, but follows a $1/\sqrt{t}$ trend at late times.}
	\vspace{-0.5cm}
	\label{EFig_mlesensitivity}
\end{figure} 

Such results are plotted in Supp. Fig.~\ref{EFig_mlesensitivity}a at a late fixed time of $t=8~\mu$s ($\Omega t/2\pi\approx42$). We see that indeed while both $F_c$ and $1-RSS$ are peaked at $\Omega/2\pi=5.3$ MHz, the $F_c$ peak is both more sharp, and has a larger dynamic range. Further, we plot the central peak width as a function of time in Supp. Fig.~\ref{EFig_mlesensitivity}b, where we see after the entangling time, $t_\mathrm{ent}$ (given as the point at which the half-chain entanglement entropy saturates), the FWHM of the $F_c$ central peak decreases as $\approx1/t$, while the FWHM of the $1-RSS$ central peak decreases as $\approx1/\sqrt{t}$, indicating that the protocol based on $F_c$ will become comparatively more sensitive as the quench time increases.

\section{Typicality of Benchmarking and associated error scaling}
Benchmarking results for RUCs are averaged over an ensemble of RUCs realized via sampling of two-qubit unitaries $\hat{U}$ (Ext.~Data Fig.~3). For $F_\text{XEB}$, a single choice of an RUC out of such an ensemble already yields accurate results in the sense that, in the limit of large circuit depth, the error from using only a single RUC scales as $1/\sqrt{D}$ with $D$ being the Hilbert space dimension of a global system~\cite{Arute2019}. We now investigate if a similar typicality scaling holds for benchmarking of Hamiltonian dynamics. To this end, we consider an ensemble of Hamiltonians generated from a mixed-field Ising model~\cite{Kim2014}, given as 
\begin{align}
\hat{H}_\text{QIMF}/\hbar = 2\pi \left[\sum_{i=1}^N 0.22 \hat{S}_i^x + 0.25 \hat{S}_i^y + \sum_{i=1}^{N-1} \hat{S}_i^x \hat{S}_{i+1}^x  \right],
\end{align}
by adding a series of site-dependent transverse fields:
\begin{align}
	\hat{H}_\text{QIMF}(\mathbf{J})/\hbar= \hat{H}_\text{QIMF}/\hbar + 2\pi\sum_{i=1}^N J_i \hat{S}_i^z, 
\end{align}
where $\hat{H}_\text{QIMF}(\mathbf{J})$ is a Hamiltonian in the ensemble parameterized by site-dependent transverse fields, $\mathbf{J}=(J_1, ...,J_i, ..., J_N)$. We sample the $J_i$ uniformly from $[-0.5,0.5]$ with a restriction, $\sum_{i=1}^N J_i=0$, imposed so all target states thermalize to effective infinite effective temperature. With the $\hat{H}_\text{QIMF}(\mathbf{J})$, we generate an ensemble of parametrized target states, $|\psi(t,\mathbf{J})\rangle = e^{-i \hat{H}_\text{QIMF}(\mathbf{J})t/\hbar} |0\rangle^{\otimes N}$. We now consider a single error occurring at a time $t_0$, induced by a single-site rotation around the $x$-axis, on top of otherwise error-free evolution. We denote the erroneous states as $|\psi(t,\mathbf{J})\rangle_e$. The rotation angle is fixed across $\mathbf{J}$-choices with an amplitude that yields a many-body overlap $F(\mathbf{J})=|\langle\psi(t,\mathbf{J})|\psi(t,\mathbf{J})\rangle_e|^2 \approx 0.5$.  

\begin{figure}[t!]
	\centering
	\includegraphics[width=89mm]{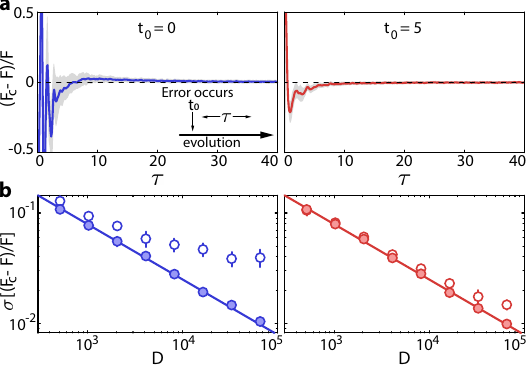}
	\caption{\textbf{Typicality analysis of $F_c$:}  \textbf{a,} Normalized fidelity estimation errors as a function of interrogation time $\tau$ after a local, instantaneous error occurs at time $t_0$ (see the inset of \textbf{a}). The error strength is chosen  such that the many-body overlap is reduced to $\sim$0.5. To investigate typical statistical errors in fidelity estimation, we simulate an ensemble of chaotic Hamiltonians at a fixed system size of $N=16$ for two error occurrence times of $t_0 = 0$ (left) and $t_0$ = 5 (right) (see Methods). The mean and standard deviation of the normalized fidelity estimation errors are plotted as the solid line and grey band, respectively. \textbf{b,} Standard deviation of normalized fidelity estimation errors, characterized at two different post-error interrogation times ($\tau = 8$; open markers, $\tau = 30$; closed markers) as a function of the total Hilbert space dimension $D$ for the two error occurrence times of $t_0 = 0$ (left) and $t_0$ = 5 (right). At the late interrogation time of $\tau = 30$, we find that the typical error from the distinct choice of a Hamiltonian follows a scaling of $\sigma[(F_c - F)/F] \sim 1/\sqrt{D}$ (solid lines), as also found for deep random unitary circuits (Methods). In \textbf{b}, error bars denote the standard deviation of $\sigma[(F_c-F)/F]$ in a $[\tau-1,\tau+1]$ window.} \vspace{-0.5cm}
	\label{EFig_typicality}
\end{figure} 

We now estimate this many-body overlap using the bitstring probabilities for these two states, yielding a time- and $\mathbf{J}$-dependent estimator $F_c(t,\mathbf{J})$. To quantify relative fluctuations, we introduce the relative difference $d(\tau,\mathbf{J})=(F_c(t_0+\tau,\mathbf{J})-F(\mathbf{J}))/F(\mathbf{J})$,  where $\tau=t-t_0$ is the evolution time after the error.  We show the ensemble average of $d(\tau,\mathbf{J})$ as a function of $\tau$ (Supp. Fig.~\ref{EFig_typicality}a) for two different choices of $t_0$: $t_0=0$ and $t_0=5$. We observe that the ensemble average settles quickly to zero within a $10^{-2}$ level; however, we find fluctuations around this mean value arising from different choices of $\mathbf{J}$. To quantify these fluctuations, we evaluate the standard deviation of $d(\tau,\mathbf{J})$ over $\mathbf{J}$ and find that it decreases with system size (Supp. Fig.~\ref{EFig_typicality}b). At late times $\tau$, we find a $1/\sqrt{D}$ scaling similar to the case of RUCs. These results suggest that, at sufficiently late time, our method becomes increasingly more precise in the limit of large system size for a single typical Hamiltonian evolution.

\section{Noisy system dynamics modeling and benchmarking of Rydberg-atom Quantum Simulator}
\label{sec:errormodel}

\textbf{Noisy system dynamics in the Rydberg model.} The chaotic dynamics of a one-dimensional Rydberg-atom array presented in this study can be described by time evolution under the time-independent Rydberg Hamiltonian, $\hat{H}$~\cite{Madjarov2019,Bernien2017}. In the ideal case without any environmental noise and control imperfections, we assume homogeneous global control and a defect-free ordered array; both Rabi frequencies~$\Omega_i = \Omega$ and detunings~$\Delta_i = \Delta$ are assumed to be site-independent and atoms are equally spaced with lattice spacing $a$. In experiments, however, quantum systems are not perfectly isolated and can undesirably interact with the randomly fluctuating environment; this causes control parameters in the Hamiltonian to drift over time and leads to a decay of quantum correlations between qubits. To this end, we design an \textit{ab initio} error model by carefully considering realistic error sources affecting our experiment. Specifically, we introduce random fluctuations in both control parameters and atomic positions such that at each site $i$, atoms experience inhomogeneous, time-dependent control field strengths as well as positional disorder, modeled as
\begin{align}
	\Omega_i &= \Omega + \delta \Omega(t) + \delta \Omega_i \\
	\Delta_i  &= \Delta + \delta \Delta(t) + \delta \Delta_i \\
	R_{ij} &= a |i-j| + \delta R_{ij}.
\end{align}
Here $\delta \Omega(t)$ and $\delta \Delta(t)$ are the \textit{time-dependent, global} fluctuations given by experimentally measured laser intensity and phase noise with no free parameters, respectively. Similarly, $\delta \Omega_i$, $\delta \Delta_i$, and $\delta R_{ij}$ are the \textit{time-independent, local} fluctuations of Rabi frequency, Doppler shifts, and atomic positions, with $i$ and $j$ denoting site indices~\cite{Madjarov2019, Madjarov2020}. In addition to these perturbations in the Hamiltonian parameters, we also take into account spontaneous decay of the Rydberg state as well as state-preparation-and-measurement errors that account for both initial loss of an atom and small infidelities in atomic state imaging and Rydberg state detection.

\begin{figure}[t!]
	\centering
	\includegraphics[width=89mm]{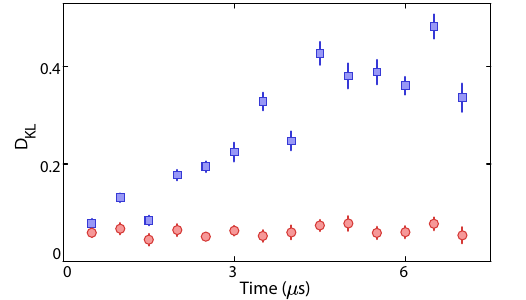}
	\caption{\textbf{Quantitative evaluation of \textbf{\textit{ab inito}} error model:} Kullback-Leibler (KL) divergence of measurement outcome probability distribution between the ten-qubit Rydberg experiment and an error-free model (blue square) and an \textit{ab initio} error model (red circle) (Methods). When the experimental data is benchmarked against an error-free model, the KL divergence increases over time since the presence of decoherence effects and control imperfections are not taken into account in the error-free model. In the case of the error model, however, we find a significant reduction in the KL divergence between the experiment and error model at \textit{all} times. Error bars denote the s.e.m.} \vspace{-0.5cm}
	\label{EFig_KLdiv}
\end{figure}

We simulate these errors in noisy quantum evolution via the Monte Carlo stochastic wavefunction method~\cite{Molmer1993}. We find that the resultant global bitstring measurement probabilities show a high degree of correlation with the experimental measurement outcome probability distribution, indicating that the error model reliably reproduces our experimental system dynamics (Supp. Fig.~\ref{EFig_KLdiv}). The accurate modeling of noisy dynamics in our system is further corroborated by good agreement with the fidelity estimator $F_c$ (Fig.~4b,d).

\textbf{Evaluation of $F_c$ in the Rydberg model.} Bitstring probability distributions in our system are characterized in a constrained Hilbert space $\mathcal{H}_c$ where the simultaneous excitation of two neighboring atoms to the Rydberg state is not allowed. In order to benchmark our quantum device in the subspace $\mathcal{H}_c$, we modify the fidelity estimator $F_c$ as
\begin{align}
	F_c(t) = B(t) B_0(t) \left( 2 \frac{\sum_{z \in \mathcal{H}_c} \tilde{p}(z) \tilde{p}_0(z)}{\sum_{z \in \mathcal{H}_c} \tilde{p}_0(z)^2} -1 \right). \label{eq:F_cRydberg}
\end{align}
Here $B(t)$ are $B_0(t)$ are the total probabilities of being in the subspace $\mathcal{H}_c$ at time $t$ in noisy and clean evolutions, respectively, and $\tilde{p}$ and $\tilde{p}_0$ are the corresponding bitstring probabilities normalized in $\mathcal{H}_c$. We numerically confirm that Eq.~(\ref{eq:F_cRydberg}) is a good approximation in the strong Rydberg blockade regime, provided that noisy evolution results in negligible many-body overlap in the manifold outside $\mathcal{H}_c$. In addition, we take into account the spatial inversion symmetry of the Rydberg Hamiltonian. Specifically, our Hamiltonian commutes with the left-right inversion operator $\hat{Q}$, which swaps two atoms at sites $i$ and $N-i+1$ for every $i$, due to the global uniformity implicit in the one-dimensional Rydberg model. Under this symmetry, we find that $\ket{z_e} = \frac{\ket{z} + \ket{\bar{z}}}{\sqrt{2}}$ and $\ket{z_o} = \frac{\ket{z} - \ket{\bar{z}}}{\sqrt{2}}$ are the even- and odd-parity eigenstates of $\hat{Q}$ with eigenvalues of $1$ and $-1$, respectively. Here, $\ket{z}$ and $\ket{\bar{z}}$ are the $z$-basis bitstring and its mirrored version, i.e., $\ket{z} = \ket{z_1 z_2 \dots z_N}$ and $\ket{\bar{z}} = \ket{z_N z_{N-1} \dots z_1}$ where $z_i$ is either 0 or 1 at site $i$. Since our initial state $\ket{\psi_0} = \ket{0}^{\otimes N}$ is the even-parity eigenstate of $\hat{Q}$, i.e., $\hat{Q} \ket{\psi_0} = +\ket{\psi_0}$, the resulting many-body states after a quench always reside within the even-parity sector, reducing the effective Hilbert space dimension approximately by a factor of two. Furthermore, since projective measurement yields either $\ket{z}$ or $\ket{\bar{z}}$ with probabilities $1/2$ due to the parity eigenbasis $\{ \ket{z_e}, \ket{z_o} \}$, their contributions to the fidelity estimation formula need to be adjusted; in consideration of this inversion symmetry, we rewrite both $\tilde{p}(z)$ and $\tilde{p}_0(z)$ using the parity eigenbasis and assume that both $\ket{z}$ and $\ket{\bar{z}}$ measurement outcomes originate from the same $\ket{z_e}$ in the even-parity sector. This effectively corresponds to checking the bitstring correlation in the even-parity sector of the parity eigenbasis. Having incorporated these modifications, we numerically confirm that the adjusted fidelity estimator shows good agreement with the true fidelity defined from the {\it full} Hilbert space (Fig.~4b,d). 

\section{Benchmarking with evolution constraints}
\begin{figure}[t!]
	\centering
	\includegraphics[width=89mm]{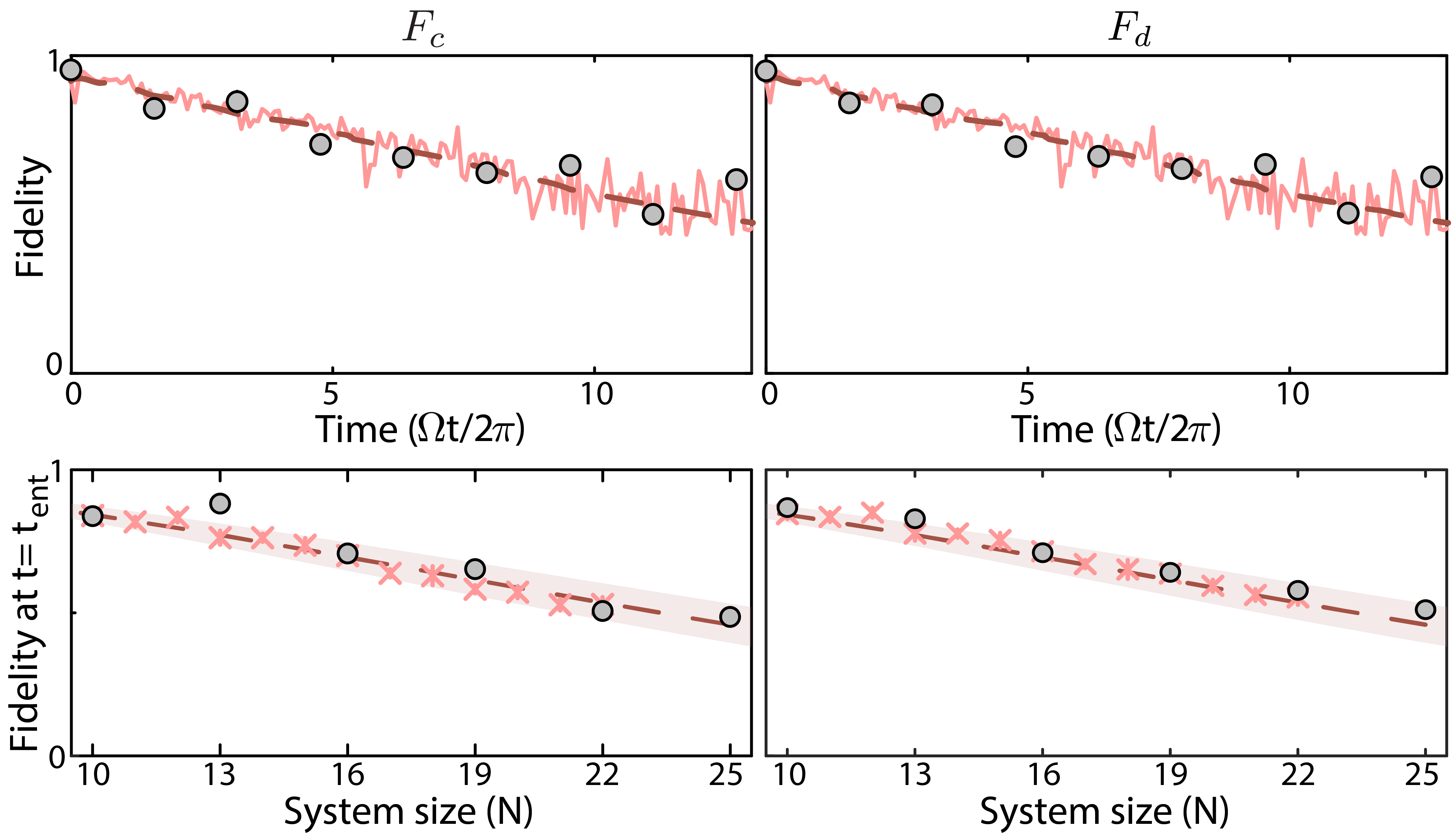}
	\caption{\textbf{Comparison to a new fidelity estimator.}  We compare benchmarking with the fidelity estimator $F_c$, as in Fig.~4 of the main text (left), against a new fidelity estimator, $F_d$~\cite{Mark2022}, which normalizes all probabilities in Eq.~2 of the main text by their time-average (right). Colors and marker types are equivalent to those in Fig.~4 in all subplots.
	} \vspace{-0.5cm}
	\label{EFig_Fd}
\end{figure}

The fidelity estimator considered in the main text, $F_c$, accurately tracks evolution fidelity when that evolution is tuned such that the initial state thermalizes at infinite effective temperature. However, as discussed in Secs.~\ref{sec:thermal} and~\ref{sec:errormodel}, problems may arise when this infinite effective temperature condition is not met, or if symmetries of the Hamiltonian restrict the system dynamics, as the projected ensemble may not completely form a 2-design. In recent work~\cite{Mark2022}, a new estimator, $F_d$, was introduced by rescaling the bitstring probabilities in Eq.~2 of the main text by their time-average; remarkably this allows for benchmarking systems evolving at finite effective temperature as well as systems of itinerant particles such as Bose- and Fermi-Hubbard models. In Supp. Fig.~\ref{EFig_Fd} we compare this new estimator, $F_d$~\cite{Mark2022}, and our original estimator $F_c$ for benchmarking our Rydberg quantum simulator, as in Fig.~4 of the main text; we observe no sizeable difference between the two estimators, which we interpet as due to intentionally performing all evolution close to infinite effective temperature in this work.

\FloatBarrier
\bibliographystyle{manubib2}
\bibliography{library_endreslab.bib}